\documentclass[aps,prx,twocolumn,superscriptaddress,nofootinbib,longbibliography]{revtex4-2}

\providecommand{\Newlabel}[2]{} 
\usepackage{amsmath,amssymb,amsfonts}
\usepackage{graphicx}
\usepackage[hidelinks]{hyperref}
\usepackage{xcolor}
\definecolor{revisionblue}{RGB}{0,70,170}
\newif\ifshowchanges
\showchangesfalse
\newcommand{\rev}[1]{{\ifshowchanges\color{revisionblue}\fi #1}}

\usepackage{amsthm}

\usepackage{enumitem}

\usepackage{setspace}

\usepackage{extarrows}

\usepackage[most]{tcolorbox}
\tcbuselibrary{breakable,skins}
\tcbset{paperboxmode/.style={unbreakable}}
\usepackage{xparse} 

\newtcolorbox{naturethmbox}[1][]{%
  enhanced,
  paperboxmode,
  sharp corners,
  boxrule=0pt,
  frame hidden,
  colback=black!4.5,
  borderline west={0.7pt}{0pt}{black!50},
  left=7pt,
  right=7pt,
  top=5pt,
  bottom=5pt,
  before skip=8pt,
  after skip=8pt,
  #1
}

\newtcolorbox{naturedefbox}[1][]{%
  enhanced,
  paperboxmode,
  sharp corners,
  boxrule=0pt,
  frame hidden,
  colback=black!4.5,
  borderline west={0.6pt}{0pt}{black!40},
  left=7pt,
  right=7pt,
  top=5pt,
  bottom=5pt,
  before skip=8pt,
  after skip=8pt,
  #1
}

\NewDocumentCommand{\bmyt}{o}{%
  \begin{naturethmbox}%
  \IfNoValueTF{#1}{\begin{theorem}}{\begin{theorem}[#1]}%
}
\newcommand{\emyt}{%
  \end{theorem}%
  \end{naturethmbox}%
}

\NewDocumentCommand{\bmyl}{o}{%
  \begin{naturethmbox}%
  \IfNoValueTF{#1}{\begin{lemma}}{\begin{lemma}[#1]}%
}
\newcommand{\emyl}{%
  \end{lemma}%
  \end{naturethmbox}%
}

\NewDocumentCommand{\bmyp}{o}{%
  \begin{naturethmbox}%
  \IfNoValueTF{#1}{\begin{proposition}}{\begin{proposition}[#1]}%
}
\newcommand{\emyp}{%
  \end{proposition}%
  \end{naturethmbox}%
}

\NewDocumentCommand{\bmyd}{o}{%
  \begin{naturedefbox}%
  \IfNoValueTF{#1}{\begin{definition}}{\begin{definition}[#1]}%
}
\newcommand{\emyd}{%
  \end{definition}%
  \end{naturedefbox}%
}

\NewDocumentCommand{\bmyc}{o}{%
  \begin{naturethmbox}%
  \IfNoValueTF{#1}{\begin{corollary}}{\begin{corollary}[#1]}%
}
\newcommand{\emyc}{%
  \end{corollary}%
  \end{naturethmbox}%
}

\usepackage{dsfont}

\usepackage{mathtools}

\def\1{\mathbf{1}}

\def\prob{{\rm Prob}}

\newcommand{\epm}{\end{pmatrix}}
\newcommand{\bpm}{\begin{pmatrix}}

\renewcommand{\log}{{\operatorname{log}}}

\newcommand{\ebm}{\end{bmatrix}}
\newcommand{\bbm}{\begin{bmatrix}}

\def\A{\boldsymbol{\mathsf{A}}}
\def\B{\boldsymbol{\mathsf{B}}}
\def\C{\boldsymbol{\mathsf{C}}}

\def\N{\mathbb{N}}

\def\X{\boldsymbol{\mathsf{X}}}
\def\Y{\boldsymbol{\mathsf{Y}}}
\def\Z{\boldsymbol{\mathsf{Z}}}

\def\>{\rangle}
\def\<{\langle}

\def\mE{\mathcal{E}}

\def\mL{\mathcal{L}}

\def\mT{\mathcal{T}}

\newcommand{\supp}{\operatorname{supp}}

\renewcommand{\ge}{\geqslant}
\renewcommand{\le}{\leqslant}

\newcommand{\ben}{\begin{enumerate}}
\newcommand{\een}{\end{enumerate}}

\theoremstyle{definition}
\newtheorem{theorem}{Theorem}
\newtheorem{lemma}[theorem]{Lemma}
\newtheorem{corollary}[theorem]{Corollary}
\newtheorem{proposition}[theorem]{Proposition}
\newtheorem{definition}[theorem]{Definition}
\newtheorem{remark}[theorem]{Remark}

\usepackage{array,tabularx,booktabs}
\let\textcaron\v 

\newcommand{\bea}{\begin{eqnarray}}
\newcommand{\eea}{\end{eqnarray}}
\newcommand{\be}{\begin{equation}}
\newcommand{\ee}{\end{equation}}
\newcommand{\ba}{\begin{equation}\begin{aligned}}
\newcommand{\ea}{\end{aligned}\end{equation}}

\newcommand{\bee}{\begin{enumerate}}
\newcommand{\eee}{\end{enumerate}}

\newtheorem{example}{Example}

\newcommand{\G}{\mathbb{G}}

\newcommand{\md}{\mathfrak{D}}

\newcommand{\mW}{\mathcal{W}}

\newcommand{\D}{\mathbb{D}}

\newcommand{\tr}{{\rm Tr}}

\newcommand{\eps}{\varepsilon}

  \def\H{\mathbb{H}}

\newcommand{\eqdef}{\coloneqq}

\def\p{\mathbf{p}}
\def\q{\mathbf{q}}

\def\t{\mathbf{t}}
\def\u{\mathbf{u}}
\def\v{\mathbf{v}}

\def\0{\mathbf{0}}

\def\tp{\tilde{\mathbf{p}}}
\def\tq{\tilde{\mathbf{q}}}

\def\d{{\mathrm{d}}}

\newcommand{\cmaj}{\succ_L^{*}}
\hypersetup{bookmarksdepth=2,pdftitle={A Single Physical Principle Determines Quantum Relative Entropy},
 pdfauthor={Gilad Gour}}
\begin{document}

\title{A Single Physical Principle Determines Quantum Relative Entropy}
\author{Gilad Gour}
\affiliation{Faculty of Mathematics, Helen Diller Quantum Center, Technion--Israel Institute of Technology, Haifa 3200003, Israel}
\affiliation{Mathematics with Computer Science Program, Guangdong Technion–Israel Institute of Technology, Shantou, Guangdong 515063, China}
\email{giladgour@technion.ac.il}

\begin{abstract}
Why does quantum relative entropy occupy such a privileged role among measures of distinguishability? We show that a single operational principle determines it. If one pair of states can be distinguished at least as successfully as another for every prior, when both are assisted by the same auxiliary pair, acting as a catalyst, a distinguishability measure should assign it no smaller value. Remarkably, for a normalized measure, this requirement singles out Umegaki relative entropy under mild classical continuity and expansibility on uniform distributions: adding impossible outcomes leaves their associated entropy unchanged. Qubit catalysts already suffice. No additivity or quantum continuity is assumed. We also prove that, without auxiliary assistance, respecting binary discrimination dominance uniquely determines the quantum extension of every regular classical divergence, including the Hirche--Tomamichel extensions of classical $f$-divergences. For additive classical divergences, these extensions respect classical catalysis; allowing noncommuting qubit catalysts is what singles out Umegaki relative entropy.
\end{abstract}

\maketitle

\section{Introduction}

Quantum relative entropy occupies a central place in quantum information theory, statistical physics and thermodynamics. It governs limits on communication, hypothesis testing and resource conversion~\cite{NC2000,Wilde2013,Watrous2018,Tomamichel2015,KW2024,Gour2025}. Its ubiquity raises a basic question: when should one pair of quantum states be considered more distinguishable than another? A measure of distinguishability should reflect such a comparison by assigning the more distinguishable pair a value at least as large. We show that comparing success in binary guessing games, with the same auxiliary pair assisting each experiment, singles out Umegaki relative entropy under mild classical regularity and entropy conventions, without assuming additivity.

There are two natural operational comparisons. The first concerns \emph{conversion}: a pair $(\rho,\sigma)$ is at least as distinguishable as $(\rho',\sigma')$ if a single quantum channel $\mE$ maps both hypotheses to their targets, $\mE(\rho)=\rho'$ and $\mE(\sigma)=\sigma'$. We call this comparison \emph{relative majorization} and write $(\rho,\sigma)\succ(\rho',\sigma')$ whenever such a channel $\mE$ exists. A measure respecting relative majorization satisfies the data-processing inequality (DPI) and is called a quantum divergence. The second concerns \emph{discrimination performance}: the source pair is at least as distinguishable as the target if it permits at least as high an optimal probability of correctly identifying the state for every prior. Each experiment uses its own optimal measurement, which may also depend on the prior.

For classical pairs, these two comparisons coincide by Blackwell's theorem~\cite{Blackwell1953}. In quantum theory, conversion still implies dominance in every binary discrimination game, but the converse generally fails~\cite{Jencova2012,BG2017,Matsumoto2014}: one pair can offer at least as good an optimal guess for every prior even though no common channel converts it into the other. Quantum theory therefore separates two meanings of distinguishability that classical theory identifies. Correspondingly, it separates the requirement that a scalar respect channel conversion from the stronger requirement that it respect all unanimous comparisons of binary discrimination performance.

\begin{figure}[tb]
\centering
\includegraphics[width=\columnwidth]{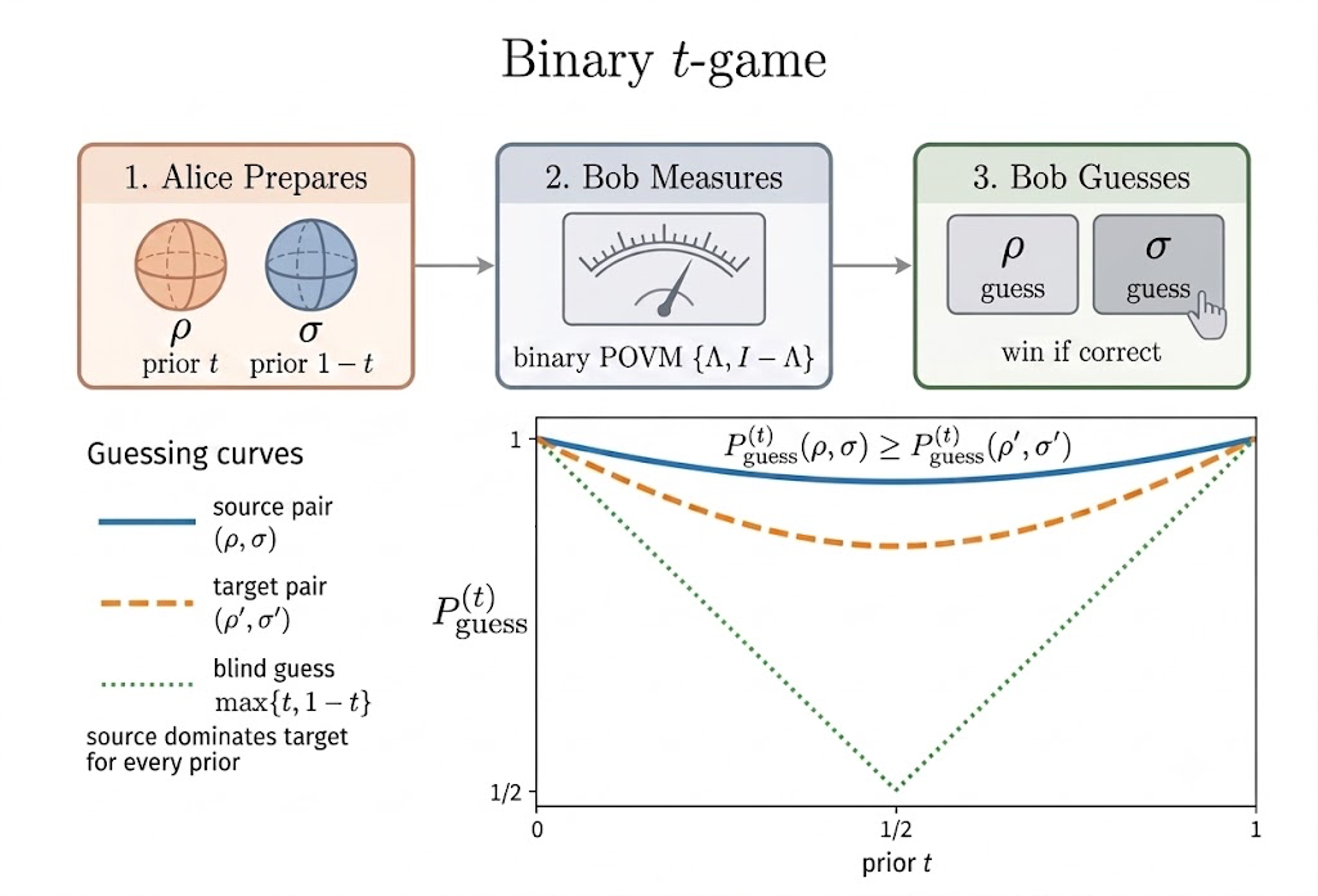}
\caption{\textbf{Lorenz majorization as dominance in binary guessing games.} In the $t$-game, Alice prepares $\rho$ with probability $t$ and $\sigma$ with probability $1-t$, and Bob performs a binary measurement to guess which state was prepared. The pair $(\rho,\sigma)$ Lorenz-majorizes $(\rho',\sigma')$ precisely when its optimal winning probability is at least as high for every $t\in[0,1]$.}
\label{tgame}
\end{figure}

Concretely, in the $t$-game (Fig.~\ref{tgame}), Alice prepares $\rho$ with prior probability $t$ and $\sigma$ with probability $1-t$. Bob receives the unknown system and performs a binary measurement to guess which state was prepared. His optimal winning probability is
\be\label{eq:binary-guessing}
 P_{\rm guess}^{(t)}(\rho,\sigma)=\max_{0\le\Lambda\le I}\bigl\{t\tr[\Lambda\rho]
+(1-t)\tr[(I-\Lambda)\sigma]\bigr\}\,.
\ee
For $t=1/2$, this is the equal-prior Helstrom discrimination problem; varying $t$ probes the pair under all possible prior odds. The source dominates the target precisely when
\be\label{t-game-dominance}
P_{\rm guess}^{(t)}(\rho,\sigma)
\ge
P_{\rm guess}^{(t)}(\rho',\sigma')
\qquad
\forall\,t\in[0,1]\,.
\ee
The same prior is used for both pairs, with each experiment using its own optimal measurement. Pairs whose winning-probability curves cross remain incomparable.

The second comparison is known as \emph{quantum Lorenz majorization}~\cite{BG2017}. We propose to adopt it as a defining principle for measures of distinguishability: whenever optimal guessing performance favors one pair for every prior, a numerical measure should preserve that ordering. We call quantum divergences satisfying this requirement \emph{quantum Lorenz divergences} (QLDs). Our central question is how strongly this operational principle constrains the measure itself. This perspective builds on the decision-theoretic comparison of quantum experiments~\cite{Shmaya2005,Buscemi2012,BDS2014,Matsumoto2015,Jencova2016} and complements approaches that quantify information through other families of discrimination tasks~\cite{TR2019,SSC2019,BGG2022,GWBG2024,GKN+2025}.

We write $(\rho,\sigma)\succ_L(\rho',\sigma')$ when~\eqref{t-game-dominance} holds. Geometrically, the binary measurements on a pair generate a testing region of accessible classical probabilities. The source Lorenz-majorizes the target when its testing region contains that of the target~\cite{BG2017}; Section~\ref{sec:binary-curves} expresses this comparison through hockey-stick curves.

\begin{figure}[htb]
\centering
\includegraphics[width=\columnwidth]{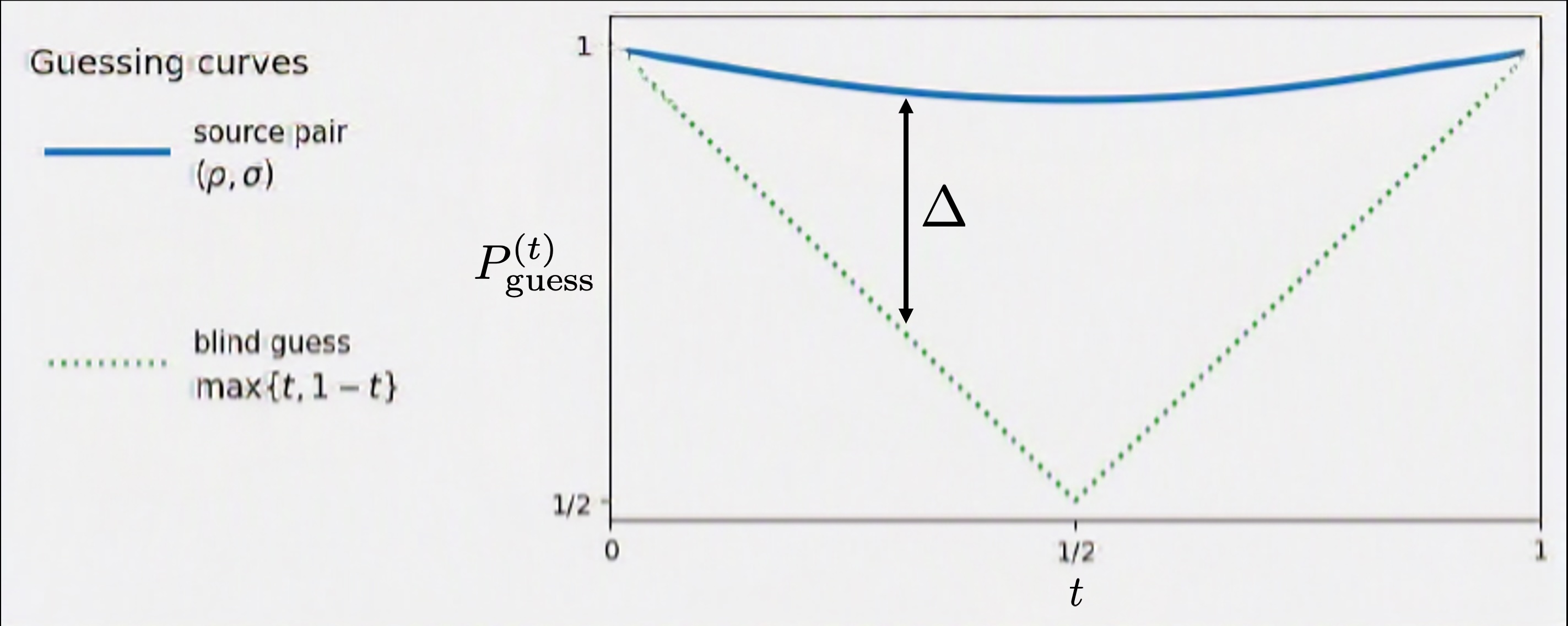}
\caption{\textbf{Advantage supplied by the experiment.} The difference $P_{\rm guess}^{(t)}(\rho,\sigma)-\max\{t,1-t\}$ measures the improvement over ignoring the system and always guessing the more likely hypothesis. Its weighted integral is relative entropy, as in~\eqref{eq:main-guessing-entropy}.}
\label{Delta}
\end{figure}

Umegaki relative entropy, 
\be
D(\rho\|\sigma)\eqdef\tr[\rho(\log\rho-\log\sigma)]\,,
\ee 
is itself a QLD. Frenkel's integral formula~\cite{Frenkel2023} represents it through binary-testing divergences. Hirche and Tomamichel~\cite[Sec.~2.5, Eq.~(2.59)]{HT2024} expressed the same identity directly in terms of the advantage in binary state discrimination:
\be\label{eq:main-guessing-entropy}
 D(\rho\|\sigma)=\frac1{\ln2}\int_0^1
 \frac{P_{\rm guess}^{(t)}(\rho,\sigma)-\max\{t,1-t\}}
 {t^2(1-t)}\,\d t\,.
\ee
We use base-two logarithms, so relative entropy is measured in bits; $\ln$ denotes the natural logarithm. The baseline $\max\{t,1-t\}$ is the winning probability obtained by ignoring the system and always guessing the more likely hypothesis. The numerator is therefore exactly the advantage supplied by access to the quantum system, illustrated in Fig.~\ref{Delta}. Relative entropy accumulates this advantage over all prior biases with positive weights, making its Lorenz monotonicity explicit.

The class of QLDs also contains many familiar operational quantities: hypothesis-testing divergences~\cite{WR2012,BD2009}, information-spectrum and max-type divergences~\cite{NH2007,DR2009,TH2013,Datta2009,DMHB2013,DL2015}, Hilbert $\alpha$-divergences~\cite{BG2017}, and the Hirche--Tomamichel (HT) $f$-divergences~\cite{HT2024}. The HT construction extends the classical hockey-stick representation of $f$-divergences~\cite{SV2016}, and has been related to and developed through quantum layer-cake representations~\cite{LHC2025,CGH+2025,CGLL2025,BCY2026}. QLDs encompass more general, possibly nonlinear functionals of the entire binary discrimination curve.

Our first main result turns these connections into an extension principle. If a classical divergence is monotone under relative majorization and Lorenz continuous, then its values uniquely fix any quantum Lorenz extension on finite-dimensional pairs satisfying $\supp(\rho)\subseteq\supp(\sigma)$. No continuity of the quantum extension is assumed. In particular, for finite convex $f$ continuous at zero, this characterizes the HT $f$-divergence by its classical restriction and the testing order, rather than by choosing an integral formula. The observation that positive hockey-stick integrals are Lorenz monotones is immediate; the converse uniqueness statement for arbitrary QLD extensions is the substantive additional conclusion. Section~\ref{sec:extension} states and proves the extension theorem.

Ordinary Lorenz monotonicity therefore fixes the quantum extension once the classical divergence is chosen. We next ask whether consistency under catalysis can determine which classical choice survives.
Catalysis provides a general way to extend both channel conversion and discrimination comparisons, familiar from resource theories of entanglement and thermodynamics~\cite{JonathanPlenio1999,Brandao2015}. A comparison that fails for the original pairs may become possible when each is combined with the same auxiliary pair. For channel conversion, for example, a common channel may map $(\rho\otimes\tau,\sigma\otimes\omega)$ to $(\rho'\otimes\tau,\sigma'\otimes\omega)$ even when no channel converts $(\rho,\sigma)$ to $(\rho',\sigma')$. We apply this same principle to discrimination performance. 

Suppose an auxiliary pair $(\tau,\omega)$ satisfying
$\supp(\tau)\subseteq\supp(\omega)$ is supplied in both experiments,
with its state determined by the same unknown hypothesis as the primary system.
This support condition excludes auxiliaries that reveal the hypothesis perfectly. If the source joint pair permits at least as high an optimal guessing probability as the target joint pair for every prior, we require the source pair itself to receive a distinguishability value no smaller than that of the target pair. The auxiliary pair $(\tau,\omega)$ is identical on both sides; only the primary pair changes. We call this requirement \emph{catalytic Lorenz monotonicity}. It is the central physical principle of the paper: it constrains which comparisons a distinguishability measure must respect, without prescribing how its values combine on tensor products.

Our second main result shows how restrictive this consistency requirement is. With mild regularity only on classical experiments, catalytic Lorenz monotonicity determines any real-valued divergence as a continuous nondecreasing function of Umegaki relative entropy. Normalization at the trivial pair and expansibility on uniform distributions then fix this function to be the identity, yielding Umegaki relative entropy itself. Expansibility requires only that the associated entropy of a uniform distribution be unchanged by adding impossible outcomes. Neither classical nor quantum additivity is assumed. More remarkably, a two-level quantum auxiliary system is sufficient: monotonicity need only hold for full-rank qubit catalyst pairs. Classical catalysts, even with arbitrary quantum source and target pairs, do not have this effect; every finite positive R\'enyi order remains possible.

Earlier characterizations of relative entropy combine DPI with assumptions on composition and regularity. Matsumoto uses tensor-power additivity and lower asymptotic semicontinuity~\cite{Matsumoto2010}; Wilming, Gallego and Eisert use continuity at a fixed full-rank reference, product additivity and super-additivity, with a thermodynamic interpretation involving correlated catalysis~\cite{WGE2017}. Here catalysis strengthens the operational comparison itself. No classical or quantum additivity, super-additivity, quantum continuity or asymptotic continuity is assumed. The regularity condition applies only to classical pairs in each bounded likelihood-ratio sector and admits every finite positive R\'enyi order. Expansibility requires only that adding impossible outcomes to a uniform distribution leave its associated entropy unchanged. The selection of Umegaki relative entropy thus comes from the comparisons supplied by noncommuting qubit catalysts. Additivity follows as a consequence of the resulting characterization.

\section{Hockey-stick curves and Lorenz monotonicity}
\label{sec:binary-curves}

We now express the operational comparison of the introduction in terms of scalar curves. We use bold sans-serif capitals $\A,\B,\C$ for quantum state pairs and $\X,\Y,\Z$ for explicitly classical pairs. Quantum pairs may have commuting states.  For $\A=(\rho,\sigma)$, we write
$P_{\rm guess}^{(t)}(\A)\eqdef P_{\rm guess}^{(t)}(\rho,\sigma)$
and $\D(\A)\eqdef\D(\rho\|\sigma)$ for any quantum divergence $\D$.

For calculations, the entire family of games is encoded by the hockey-stick curve
\be\label{eq:main-hockey-stick}
 h_{\A}(\gamma)\eqdef E_\gamma(\rho\|\sigma)
 \eqdef\tr(\rho-\gamma\sigma)_+\,,
 \qquad\gamma\ge0\,,
\ee
where $+$ denotes the positive part. For $0<t<1$,
\be\label{eq:main-guessing-curve}
 P_{\rm guess}^{(t)}(\A)
 =1-t+t\,h_{\A}\bigl((1-t)/t\bigr)\,.
\ee
Consequently, Lorenz majorization is simply domination of scalar functions:
\be\label{eq:main-lorenz-curves}
 \A\succ_L\B
 \quad\Longleftrightarrow\quad
 h_{\A}(\gamma)\ge h_{\B}(\gamma)
 \quad\forall\,\gamma\ge0\,.
\ee
Equivalently, the testing region
\be\label{eq:main-testing-region}
 \mT(\rho,\sigma)
 =\bigl\{(\tr[\Lambda\sigma],\tr[\Lambda\rho]):0\le\Lambda\le I\bigr\}
\ee
contains that of the target~\cite{BG2017}. We work primarily with hockey-stick curves; the equivalent Lorenz-boundary and hypothesis-testing descriptions, and their geometry, are recalled in Appendices~\ref{supp:equiv-lorenz} and~\ref{supp:testing-geometry}.

Unless a general-support statement is explicitly made, all pairs are finite-dimensional and satisfy $\supp\rho\subseteq\supp\sigma$, including catalyst pairs. Equivalently, $\rho\le R\sigma$ for some finite $R$. Their hockey-stick curves satisfy the following definition. 

\bmyd\label{def:cat-admissible}
A function $h:[0,\infty)\to[0,1]$ is \emph{admissible} if it is convex, nonincreasing and $1$-Lipschitz, with $h(0)=1$ and $h(\gamma)=0$ for all $\gamma\ge R$, for some finite $R\ge1$.
\emyd

Every admissible curve is continuous by the Lipschitz condition, and the hockey-stick curve of any pair of states $(\rho,\sigma)$ with $\supp(\rho)\subseteq\supp(\sigma)$ is admissible. The trivial pair $(\rho,\rho)$ has curve $e(\gamma)=(1-\gamma)_+$: the hypotheses are identical, so no measurement distinguishes them. Every admissible curve satisfies $h(\gamma)\ge\max\{h(0)-\gamma,0\}=e(\gamma)$ by nonnegativity and the $1$-Lipschitz condition.

The proximity of two admissible curves $h$ and $k$ is measured by the uniform norm
\be\label{eq:main-uniform-norm}
 \|h-k\|_\infty\eqdef\sup_{\gamma\ge0}|h(\gamma)-k(\gamma)|.
\ee
This is the largest vertical gap between the curves, also known as the stop-loss distance~\cite{KV2022,DDGK2005}. We use only the uniform-norm notation and call convergence in this norm \emph{uniform convergence}. Appendix~\ref{supp:curve-distance} relates it to the familiar Lorenz-boundary picture retained in some figures.

\bmyd\label{def:QLD}
A functional $\D$ is a quantum Lorenz divergence (QLD) if
\be
 \A\succ_L\B\quad\Longrightarrow\quad\D(\A)\ge\D(\B)\,.
\ee
\emyd

For qubit-to-qubit pairs, Lorenz majorization is equivalent to channel conversion by the Alberti--Uhlmann theorem~\cite{AU1980}; extensions and higher-dimensional failures are discussed in~\cite{DBS2020,Matsumoto2014}. Classically the order is relative majorization, also studied as $d$-majorization, matrix majorization and thermo-majorization~\cite{Veinott1971,Dahl1999,OH2013}.

In this notation, the guessing-advantage identity~\eqref{eq:main-guessing-entropy} takes Frenkel's hockey-stick form~\cite{Frenkel2023,HT2024}:
\be\label{eq:main-entropy-integral}
 D(\rho\|\sigma)
 =\frac1{\ln2}\int_0^\infty
       \frac{h_{\A}(\gamma)-e(\gamma)}\gamma\,\d\gamma\,.
\ee
We now use the curve formulation to determine quantum divergences from their classical restrictions.

\section{Classical values fix the quantum extension}
\label{sec:extension}

For a classical pair $\X=(\p,\q)$,
\be\label{eq:main-classical-curve}
 h_{\X}(\gamma)=\sum_x(p_x-\gamma q_x)_+\,.
\ee
The regularity needed here concerns only classical experiments with bounded evidence. For $R\ge1$, let
\be\label{loc}
 \mathsf L_R\eqdef\{(\p,\q):\p,\q\in\prob(d),\ d\in\N,\ \p\le R\q\}.
\ee
The alphabet size is unrestricted. A pair on a finite alphabet belongs to some $\mathsf L_R$ exactly when $\supp(\p)\subseteq\supp(\q)$; we call such a pair, and its hockey-stick curve, \emph{finite classical}. The bound $\p\le R\q$ means that $h_{\X}$ vanishes for $\gamma\ge R$, so the curve is admissible. Finite classical curves are piecewise linear and can bracket every admissible curve, as proved in Appendix~\ref{supp:classical-brackets}. Their proximity is measured by~\eqref{eq:main-uniform-norm}.
 
 \bmyd\label{def:classical-lorenz-continuity}
A classical divergence is \emph{Lorenz continuous} if, for each finite $R\ge 1$, it is uniformly continuous on $\mathsf L_R$ in the norm~\eqref{eq:main-uniform-norm}.
\emyd

The continuity modulus may depend on $R$; no quantum continuity, asymptotic continuity or uniform control as $R\to\infty$ is required. Operationally, the condition says that classical experiments with nearly identical optimal success probabilities receive nearly identical values. Indeed, if
\be\label{eq:guessing-distance}
 d_{\rm guess}\eqdef\sup_{0\le t\le1}
 \bigl|P_{\rm guess}^{(t)}(\X)-P_{\rm guess}^{(t)}(\Y)\bigr|,
\ee
then on $\mathsf L_R$, equation~\eqref{eq:main-guessing-curve} gives
\be\label{eq:guessing-distance-comparison}
 d_{\rm guess}\le\|h_{\X}-h_{\Y}\|_\infty
 \le(1+R)d_{\rm guess}.
\ee
Every finite positive classical R\'enyi order satisfies this condition, as does every fixed probability mixture of these orders. Thus regularity does not distinguish order one. The sharp endpoints $D_{\min}$ and $D_{\max}$ fail the condition; Appendix~\ref{supp:lorenz-continuity} supplies the proofs and explains why an upper-threshold bound is needed.

Let $\D_{\rm cl}$ be monotone under classical relative majorization. Following the extremal-extension viewpoint of~\cite{GT2020}, define
\ba\label{eq:main-envelopes}
 \underline\D(\A)&=\sup_{\X:\,\A\succ_L\X}\D_{\rm cl}(\X),\\
 \overline\D(\A)&=\inf_{\X:\,\X\succ_L\A}\D_{\rm cl}(\X),
\ea
where $\X$ ranges over finite classical pairs. Every QLD extending $\D_{\rm cl}$ lies between these envelopes.

\bmyt\label{unique-lorenz-extension}
Every classical divergence that is monotone under relative majorization and Lorenz continuous has a unique QLD extension to finite-dimensional pairs with $\supp\rho\subseteq\supp\sigma$. It is given by
\be
 \underline\D(\rho\|\sigma)=\overline\D(\rho\|\sigma).
\ee
\emyt

\begin{proof}[Proof idea]
The classical brackets of Appendix~\ref{supp:classical-brackets} satisfy
\be\label{eq:main-classical-brackets}
 h_{\X_n^-}\le h_{\A}\le h_{\X_n^+},
 \qquad\|h_{\X_n^+}-h_{\X_n^-}\|_\infty\longrightarrow0,
\ee
within one fixed upper-threshold sector. Classical uniform continuity makes the bracket values Cauchy and forces their gap to zero. Lorenz monotonicity squeezes every extension to their common limit. The same construction defines the extension, and shows that it respects pointwise order. No continuity of the quantum functional is assumed. The full proof, including the completion argument on all admissible curves, is given in Appendix~\ref{app:curve-completion}.
\end{proof}

For finite convex $f:[0,\infty)\to\mathbb R$, continuous at zero and with $f(1)=0$, the classical $f$-divergence $\sum_xq_x f(p_x/q_x)$ is Lorenz continuous. Its unique QLD extension is therefore the HT $f$-divergence. For twice differentiable $f$, this takes the particularly simple form
\be\label{eq:main-ht-f}
 D_f(\A)=\int_0^\infty f''(\gamma)
              \bigl(h_{\A}(\gamma)-e(\gamma)\bigr)\,\d\gamma.
\ee
The classical representation is due to Sason and Verd\'u~\cite{SV2016}; the quantum construction is due to Hirche and Tomamichel~\cite{HT2024}, with layer-cake formulations developed in~\cite{LHC2025,CGH+2025,CGLL2025,BCY2026}. The units in~\eqref{eq:main-ht-f} are determined by $f$: $f(r)=r\log_2(r)$ gives relative entropy in bits. Appendix~\ref{supp:lorenz-convex-order} recalls the equivalent likelihood-ratio measure representation and convex order~\cite{SS2007}; no measure representation is needed for our main proof.

More generally, Theorem~\ref{unique-lorenz-extension} gives
\be\label{eq:main-curve-functional}
 \D(\A)=\G[h_{\A}],
\ee
where $\G$ is pointwise monotone and uniformly continuous on each fixed upper-threshold sector of admissible curves. It need not be affine. For example, the HT R\'enyi divergences are obtained from
\ba\label{eq:main-renyi-curves}
 Q_\alpha[h]&=1+\alpha(\alpha-1)\int_0^\infty
          \gamma^{\alpha-2}(h(\gamma)-e(\gamma))\,\d\gamma,\\
 D_\alpha(\A)&=\frac{\log Q_\alpha[h_{\A}]}{\alpha-1},
 \qquad 0<\alpha<\infty,\quad\alpha\ne1,
\ea
with $D_1=D$. The extension principle determines these quantum quantities from their classical values and the binary comparison alone.

\section{Catalytic comparison selects relative entropy}
\label{sec:catalytic-main}

To compare $\A=(\rho,\sigma)$ with $\B=(\rho',\sigma')$, supply both experiments with the same auxiliary pair $\C=(\tau,\omega)$, satisfying $\supp(\tau)\subseteq\supp(\omega)$. Under the first hypothesis the auxiliary state is $\tau$; under the second it is $\omega$. Bob therefore distinguishes $\rho\otimes\tau$ from $\sigma\otimes\omega$ in the first experiment, and $\rho'\otimes\tau$ from $\sigma'\otimes\omega$ in the second. Both experiments use the same prior. The primary and auxiliary systems are independent conditional on the hypothesis, and Bob optimizes a joint measurement separately for each experiment.

Our principle is that if the common auxiliary pair makes the first experiment at least as successful as the second for every prior, then $\D(\A)\ge\D(\B)$. The conclusion concerns the original pairs: the same auxiliary information is available on both sides and should not reverse their ranking by the divergence. This requirement constrains the ordering of divergence values without assuming additivity on tensor products. We formalize it as follows.

\bmyd\label{def:cat-order}
We write $\A\cmaj\B$ if some finite-dimensional pair $\C=(\tau,\omega)$ with $\supp(\tau)\subseteq\supp(\omega)$ satisfies
\be\label{eq:cat-order}
 \A\otimes\C\succ_L\B\otimes\C\,.
\ee
A functional $\D$ is catalytically Lorenz monotone if
\be\label{eq:main-catalytic-monotonicity}
 \A\cmaj\B\quad\Longrightarrow\quad\D(\A)\ge\D(\B).
\ee
\emyd

Equivalently, for that same $\C$ in~\eqref{eq:cat-order},
\be\label{eq:main-catalytic-guessing}
 P_{\rm guess}^{(t)}(\A\otimes\C)\ge
 P_{\rm guess}^{(t)}(\B\otimes\C)\qquad\forall\,t\in[0,1]\,.
\ee
The catalyst may depend on the comparison, but one catalyst must work for all priors in~\eqref{eq:main-catalytic-guessing}. 
The three relations satisfy
\be\label{eq:comparison-hierarchy}
 \A\succ\B
 \quad\Longrightarrow\quad
 \A\succ_L\B
 \quad\Longrightarrow\quad
 \A\cmaj\B,
\ee
where $\A\succ\B$ denotes quantum relative majorization: a common channel converts both states of $\A$ into the corresponding states of $\B$. The first implication follows because a channel cannot improve optimal discrimination performance. The second follows by choosing an uninformative full-rank qubit catalyst $\C=(\tau,\tau)$, which leaves every optimal guessing probability unchanged. Consequently, every functional monotone under catalytic Lorenz majorization is a QLD and, in particular, a quantum divergence satisfying DPI.

The auxiliary pair is held fixed \emph{between the two experiments}; the definition concerns optimal joint discrimination performance, rather than a channel that converts both hypotheses or returns the auxiliary state after measurement.

\subsection{Associated entropy and unused dimensions.}
Following~\cite{GT2020}, we say that a quantum divergence $\D$ (i.e., functional satisfying DPI) is normalized if at the trivial system:
\be\label{eq:main-zero-normalization}
 \D(1\|1)=0,
\ee
where $1$ is the density matrix of the one-dimensional system. Discarding the system maps any pair to $(1,1)$, so DPI and $\D(1\|1)=0$ give $\D(\rho\|\sigma)\ge0$. Conversely, preparing $\rho$ maps $(1,1)$ to $(\rho,\rho)$, giving $\D(\rho\|\rho)\le0$ and hence $\D(\rho\|\rho)=0$.

For every normalized \emph{classical} divergence $\D$, define its associated entropy $\H$ on a probability vector $\p$ with $d$ components by
\be\label{eq:main-associated-entropy}
 \H(\p)\eqdef\log(d)-\D(\p\|\mathbf u_d),
\ee
where $\mathbf u_d\eqdef(1/d,\ldots,1/d)$ is the uniform distribution of the same dimension as $\p$. Both $\D$ and $\H$ accept inputs of arbitrary finite dimension. Nonnegativity of $\D$ and its vanishing on identical distributions give
\be\label{eq:main-entropy-maximum}
 \H(\p)\le\H(\mathbf u_d)=\log(d).
\ee

The remaining convention concerns the description of the same uncertainty in a larger alphabet. Expansibility is the familiar entropy requirement that adding zero-probability outcomes changes no entropy~\cite{Khinchin1957}. We need it only for uniform distributions.

\bmyd\label{def:uniform-expansibility}
The associated entropy satisfies \emph{expansibility on uniform distributions} if
\be\label{eq:main-uniform-expansibility}
 \H(\mathbf u_k\oplus\mathbf0_{m})=\H(\mathbf u_k),
 \qquad\forall\, k,m\in\N\,,
\ee
where $\mathbf0_{m}$ denotes $m$ appended zero entries.
\emyd

By~\eqref{eq:main-entropy-maximum}, $\H(\mathbf u_k)=\log(k)$, so~\eqref{eq:main-uniform-expansibility} equivalently requires $\H(\mathbf u_k\oplus\mathbf0_{m})=\log(k)$. Thus the entropy of $k$ equally likely outcomes is independent of how many impossible outcomes are listed. In quantum language, the same diagonal state is maximally mixed on a $k$-dimensional support embedded in a larger space. The condition concerns this elementary family only. Every classical R\'enyi entropy of positive order satisfies the stronger condition of expansibility for arbitrary distributions.

\bmyt\label{thm:cat-uniqueness}\label{umegaki-uniqueness}
Let $\D$ be a real-valued functional on finite-dimensional pairs with $\supp\rho\subseteq\supp\sigma$, normalized by $\D(1\|1)=0$, with Lorenz-continuous classical restriction. Suppose its associated entropy~\eqref{eq:main-associated-entropy} satisfies expansibility on uniform distributions. If $\D$ is catalytically Lorenz monotone, then
\be\label{eq:main-catalytic-uniqueness}
 \D(\rho\|\sigma)=D(\rho\|\sigma).
\ee
\emyt
\begin{remark}
It suffices to impose monotonicity for comparisons with full-rank qubit catalyst pairs.
\end{remark}

The proof establishes a more general structural statement. Without the normalization and expansibility conditions, catalytic Lorenz monotonicity and classical Lorenz continuity imply
\be\label{eq:main-catalytic-representation}
 \D(\rho\|\sigma)=g\bigl(D(\rho\|\sigma)\bigr)
\ee
for a continuous nondecreasing $g:[0,\infty)\to\mathbb R$. Conversely, every such function of Umegaki relative entropy satisfies these monotonicity and regularity conditions. Indeed, relative entropy is Lorenz monotone by~\eqref{eq:main-entropy-integral}, and its tensor-product additivity cancels the common finite catalyst contribution. This uses a property of the known entropy, not an assumption on $\D$.

The physical principle therefore fixes the ordering of entropy values, while normalization and expansibility on uniform distributions fix their numerical scale. Appendix~\ref{app:expansibility} derives the elementary consequences of these conditions; Appendix~\ref{app:main-proof} uses them to fix $g$. No additivity, super-additivity, quantum continuity or asymptotic continuity of the unknown functional is assumed. The auxiliary dimension does not grow: the proof needs only full-rank two-level pairs, including arbitrarily small noncommuting perturbations of a fixed reference state. Qubits are the smallest systems that can supply such noncommutativity.

\subsection{Why the catalytic constraint is decisive.}
The extension theorem first reduces the unknown divergence to a functional $\G$ of an admissible curve. On these curves there is a classical product $\odot$: when $\X=(\p,\q)$ is classical,
\be\label{eq:main-mixed-product}
 (h_{\A}\odot h_{\X})(\gamma)
 =\sum_{x:p_x>0}p_xh_{\A}(\gamma q_x/p_x)
 =h_{\A\otimes\X}(\gamma).
\ee
Uniform approximation extends this operation to all admissible curves. For two quantum pairs, however, $h_{\A}\odot h_{\C}$ can differ from the actual joint curve $h_{\A\otimes\C}$. Still, catalytic Lorenz monotonicity, combined with classical brackets, forces
\be\label{eq:main-replacement-principle}
\G[h_{\A}\odot h_{\C}]=\G[h_{\A\otimes\C}].
\ee
The equality also holds when the two curves appear with the same weight in an arbitrary common convex mixture; Lemma~\ref{lem:cat-replacement} proves this without assuming affinity of $\G$.

The proof uses the differences between these two product curves to generate changes that preserve $\G$. Small noncommuting qubit pairs provide sufficiently many such changes to connect, by uniform approximation, any two curves with the same relative entropy. Consequently $\G$ can depend only on the entropy integral~\eqref{eq:main-entropy-integral}, giving~\eqref{eq:main-catalytic-representation}. The full argument, including the explicit qubit calculation, is in Appendix~\ref{app:main-proof}. It is an exact finite-dimensional characterization; limiting approximations enter the proof, not the physical axiom.

\bmyc\label{cor:additive-umegaki}
Umegaki relative entropy is the only QLD with Lorenz-continuous classical restriction that is additive on independent pairs and normalized by $\D((1,0)\|(1/2,1/2))=1$.
\emyc

\begin{proof}
If $\A\otimes\C\succ_L\B\otimes\C$, Lorenz monotonicity and additivity give
\[
 \D(\A)+\D(\C)\ge\D(\B)+\D(\C).
\]
Thus $\D$ is catalytically Lorenz monotone. Additivity also gives $\D(1\|1)=0$. Lemma~\ref{lem:app-additive-expansibility} shows that the one-bit normalization and classical additivity imply expansibility on uniform distributions. Theorem~\ref{thm:cat-uniqueness} therefore gives $\D=D$.
\end{proof}

Classical additivity permits a broad family. For classical experiments with likelihood ratios bounded away from zero and infinity, Mu, Pomatto, Strack and Tamuz~\cite[Theorem~2]{MPST2021} characterize additive divergences satisfying DPI and finite on this domain as positive integrals of forward and reverse R\'enyi divergences. The catalytic theorem does not require this classification or a composition axiom for the unknown quantum functional.

\section{Why the catalyst must be quantum}
\label{sec:classical-catalysis}

The distinction between classical and quantum aids identifies the mechanism behind the theorem. A classical catalyst is a pair of probability distributions; any commuting quantum catalyst is equivalent to one by simultaneous diagonalization. The primary source and target may still be arbitrary quantum pairs.

\bmyp\label{lem:cat-classical-additivity}
Let $\D$ be a  QLD with Lorenz-continuous, additive classical restriction. Then, for every quantum pair $\A$ and classical pair $\Z$,
\be\label{eq:cat-classical-additivity}
 \D(\A\otimes\Z)=\D(\A)+\D(\Z).
\ee
Hence $\D$ respects catalytic Lorenz comparisons with classical catalysts. It also respects comparisons with arbitrary quantum catalysts when both source and target are classical.
\emyp

For a classical catalyst $\Z$, equation~\eqref{eq:cat-classical-additivity} gives
\[
\begin{gathered}
 \A\otimes\Z\succ_L\B\otimes\Z\\
 \Longrightarrow\quad
 \D(\A)+\D(\Z)\ge\D(\B)+\D(\Z)\\
 \Longrightarrow\quad
 \D(\A)\ge\D(\B).
\end{gathered}
\]
Thus the catalyst's contribution cancels. Appendix~\ref{app:classical-catalysis} derives the additivity identity from Theorem~\ref{unique-lorenz-extension} and the mixed-product formula~\eqref{eq:main-mixed-product}. In particular, for every $0<\alpha<\infty$, the HT $\alpha$-R\'enyi divergence satisfies
\be\label{eq:cat-renyi-classical-additivity}
 D_\alpha(\A\otimes\Z)=D_\alpha(\A)+D_\alpha(\Z),
\ee
and is therefore monotone under classical catalysis.

For every finite $\alpha>0$, $D_\alpha(1\|1)=0$ and the classical restriction of $D_\alpha$ is Lorenz continuous. Its associated classical entropy is the R\'enyi entropy $H_\alpha$, which satisfies the stronger expansibility condition
\[
 H_\alpha(\p\oplus\mathbf0_m)=H_\alpha(\p)
\]
for every probability vector $\p$. Fixed probability mixtures of these divergences preserve the normalization, classical Lorenz continuity and expansibility, and remain monotone under classical catalysis.

Classical catalysts therefore do not single out relative entropy: every finite positive R\'enyi order remains compatible with the same normalization, expansibility and classical continuity requirements. Proposition~\ref{lem:cat-classical-additivity} also shows that allowing quantum catalysts is insufficient if both primary pairs are restricted to be classical. Uniqueness requires allowing noncommutativity in both the catalyst and the pairs being compared. However, the two primary pairs need not both be noncommuting: our proof uses only catalytic comparisons with at least one classical endpoint, as explained in Appendix~\ref{app:classical-catalysis}.

Classical catalytic monotonicity is also a genuine constraint beyond ordinary Lorenz monotonicity. It is not automatically satisfied by every HT $f$-divergence.

\begin{example}[Failure of classical catalytic monotonicity]\label{ex:cat-ht-counterexample}
The smooth strictly convex function
\be\label{eq:main-smooth-f}
 f(t)=\sqrt{(t-1)^2+1/16}-1/4
\ee
defines a Lorenz-continuous HT divergence that fails monotonicity under classical catalysis. Using the Jonathan--Plenio catalyst~\cite{JonathanPlenio1999}, one obtains classical pairs $\X,\Y,\Z$ satisfying
\be
 \begin{gathered}
 \X\otimes\Z\succ_L\Y\otimes\Z,\\
 D_f(\X)=\frac{\sqrt{17}-1}{8}<\frac25=D_f(\Y).
 \end{gathered}
\ee
The explicit pairs and the verification are given in Appendix~\ref{app:classical-catalysis}.
\end{example}

Thus classical aids eliminate some QLDs while preserving the R\'enyi family; noncommuting qubit aids impose the additional comparisons that select relative entropy. Approximating a quantum catalyst's individual curve by classical curves does not reproduce its effect in a joint experiment: the approximation yields $h_{\A}\odot h_{\C}$, which need not equal $h_{\A\otimes\C}$.

\section{Discussion}

The choice of a distinguishability measure begins with the comparisons it is intended to represent. Channel conversion and binary discrimination dominance coincide classically and separate quantumly. Catalysis provides a general way to extend either comparison by supplying the same auxiliary pair on both sides. Applied to discrimination performance, it requires a scalar to respect unanimous optimal guessing comparisons made with this common aid.

Two results establish the consequences. Ordinary Lorenz monotonicity makes the quantum extension of a classically regular divergence unique. Catalytic Lorenz monotonicity then removes the freedom to choose the underlying entropy, leaving only a continuous nondecreasing function of Umegaki relative entropy. Normalization at the trivial pair and expansibility on uniform distributions fix its values in bits. The latter requires the associated entropy to be independent of unused dimensions, even just for uniform distributions. Additivity then emerges as a conclusion rather than serving as a premise. The quantum auxiliary systems needed to obtain this conclusion are full-rank qubits, not catalysts of unbounded dimension.

The role of regularity is limited but necessary. The additive QLDs $D_{\min}$ and $D_{\max}$ satisfy the normalization, expansibility and catalytic monotonicity conditions, yet their classical restrictions fail Lorenz continuity. These endpoints show why a robustness condition cannot simply be omitted. Every finite positive R\'enyi order satisfies that same robustness condition and expansibility on uniform distributions, and survives all classical catalytic comparisons. It is the additional comparison supplied by noncommuting quantum aids that rules out the other orders.

Theorem~\ref{thm:cat-uniqueness} also shows how an alternative measure must fail the proposed principle. Let $\D$ be a QLD satisfying normalization~\eqref{eq:main-zero-normalization}, expansibility on uniform distributions and classical Lorenz continuity. If $\D\ne D$, there exist primary pairs $\A,\B$ and a full-rank qubit catalyst $\C$ such that~\eqref{eq:main-catalytic-guessing} holds and yet $\D(\A)<\D(\B)$.
Thus the measure assigns a smaller value to $\A$ even though, with the same auxiliary pair, it can be distinguished at least as successfully as $\B$ for every prior. The theorem guarantees that such a comparison exists; finding explicit pairs for a given alternative divergence is a separate problem. This obstruction concerns consistency with assisted binary discrimination, while other divergences may quantify different operational tasks.

The connection to catalytic thermodynamics is a motivation for examining common resources, rather than an identification of different operational theories. Catalytic Lorenz majorization compares testing performance, whereas a thermodynamic process specifies physical dynamics and restoration of an ancillary state~\cite{Brandao2015}. Understanding which implementations, restrictions on measurements or repeatable auxiliary uses lead to related uniqueness principles is an interesting further problem. Infinite-dimensional systems and unbounded evidence require additional domain and continuity analysis. Multi-hypothesis experiments offer another direction, where binary curves are replaced by richer decision regions. These questions extend the operational comparison approach beyond the finite-dimensional binary setting considered here.

\begingroup
\setlength{\parindent}{0pt}
\setlength{\parskip}{0pt}

\noindent\textbf{Data availability.}
No datasets were generated or analysed in this theoretical study.
\par\smallskip

\noindent\textbf{Funding.}
This research was supported by the Israel Science Foundation
under Grant No.~1192/24.
\par\smallskip

\noindent\textbf{Acknowledgements.}
The author thanks Nilanjana Datta and Goni Yoeli for
stimulating discussions.
\par\smallskip

\noindent\textbf{Use of artificial intelligence.}
The author used GPT-6 Astra via Codex (OpenAI) to assist
with mathematical exploration and manuscript preparation.
The author verified the mathematical arguments and reviewed
and revised the manuscript, and takes full responsibility
for its content.
\par\smallskip

\noindent\textbf{Author contributions.}
The author conceived the study, developed the theoretical
framework, derived and proved the results, and wrote and
revised the manuscript.
\par\smallskip

\noindent\textbf{Competing interests.}
The author declares no competing interests.
\par

\endgroup

\onecolumngrid
\appendix
\tcbset{paperboxmode/.style={breakable}}
\setcounter{section}{0}
\setcounter{equation}{0}
\setcounter{figure}{0}
\renewcommand{\theequation}{\thesection.\arabic{equation}}
\renewcommand{\thefigure}{\thesection.\arabic{figure}}
\numberwithin{equation}{section}
\numberwithin{figure}{section}
\renewcommand{\theHequation}{appendix.\thesection.\arabic{equation}}
\renewcommand{\theHfigure}{appendix.\thesection.\arabic{figure}}

\section*{Appendix}

The appendices are arranged in proof order: classical approximation and curve completion, classical continuity, and consequences of expansibility precede the proof of Theorem~\ref{thm:cat-uniqueness}. The remaining appendices give the classical-catalysis results and the equivalent geometric and measure descriptions.

\section{Classical approximation and curve completion}
\label{supp:classical-brackets}
\label{supp:curve-completion}

We first establish the approximation results needed for Theorem~\ref{unique-lorenz-extension} and the catalytic proof. Admissibility and the uniform norm are as in Definition~\ref{def:cat-admissible} and~\eqref{eq:main-uniform-norm}. Finite classical pairs have the support condition specified in Section~\ref{sec:extension}: $\supp(\p)\subseteq\supp(\q)$. Coordinates with $p_x=q_x=0$ may be discarded.

\subsection{Finite classical representatives}

For a finite classical pair $\X=(\p,\q)$, its \emph{likelihood ratios} are $p_x/q_x$ for $q_x>0$. Equation~\eqref{eq:main-classical-curve} becomes
\be\label{eq:app-likelihood-curve}
 h_{\X}(\gamma)=\sum_{x:q_x>0}q_x\left(\frac{p_x}{q_x}-\gamma\right)_+.
\ee
Thus $p_x/q_x\le R$ for all such $x$ is equivalent to $h_{\X}(\gamma)=0$ for $\gamma\ge R$. Likewise, the ratios are \emph{bounded away from zero}, $p_x/q_x\ge\delta>0$, exactly when $h_{\X}(\gamma)=1-\gamma$ on $[0,\delta]$: no negative term is discarded on that interval. For general admissible curves, these initial and final curve conditions are used directly, without requiring a finite classical representation.

Conversely, every piecewise-linear admissible curve $k$ is finite classical. To see this, let $t_1,\ldots,t_m\in(0,R]$ be its positive breakpoints, including its final change of slope to zero, and set
\be\label{eq:app-polygonal-weights}
 q_x=k'_+(t_x)-k'_-(t_x),\quad p_x=t_xq_x\quad(1\le x\le m),
 \qquad q_0=1+k'_+(0),\quad p_0=0.
\ee
Here $k'_-$ and $k'_+$ denote the slopes immediately to the left and right of a breakpoint. Convexity gives $q_x\ge0$, and the $1$-Lipschitz condition gives $q_0\ge0$. The slope jumps telescope to $\sum_{x=1}^m q_x=-k'_+(0)$, so $\sum_{x=0}^m q_x=1$. Moreover,
\be\label{eq:app-polygonal-realization}
 k(\gamma)=\sum_{x=1}^m q_x(t_x-\gamma)_+
 =\sum_{x=0}^m(p_x-\gamma q_x)_+.
\ee
The first identity follows because both sides have the same slopes and vanish for $\gamma\ge R$. Evaluating at zero gives $\sum_xp_x=k(0)=1$. Thus $\X=(\p,\q)\in\mathsf L_R$ and $k=h_{\X}$. If $k=1-\gamma$ on $[0,\delta]$, then $q_0=0$ and all nonzero slope jumps occur at $t_x\ge\delta$, so every retained likelihood ratio belongs to $[\delta,R]$.

\subsection{Uniform classical brackets}

\bmyl[Classical brackets for admissible curves]\label{lem:bounded-classical-brackets}
Let $h$ be admissible and vanish for $\gamma\ge R$. There exist finite classical pairs $\X_n^-,\X_n^+\in\mathsf L_R$ such that
\be\label{eq:app-admissible-brackets}
 h_{\X_n^-}\le h\le h_{\X_n^+},
 \qquad\|h_{\X_n^+}-h_{\X_n^-}\|_\infty\longrightarrow0.
\ee
If, in addition, $h(\gamma)=1-\gamma$ on $[0,\delta]$ for some $0<\delta<1$, both brackets can be chosen to agree with that segment. Their likelihood ratios then all belong to $[\delta,R]$.
\emyl

\begin{figure}[ht]
\centering
\includegraphics[width=0.85\textwidth]{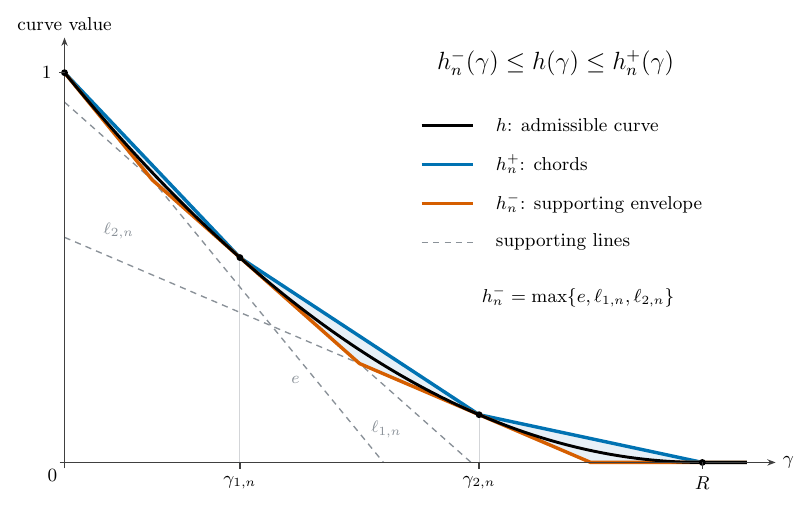}
\caption{Classical brackets for an admissible hockey-stick curve $h$ (black). The upper bracket $h_n^+$ (blue) joins the values of $h$ at the partition points by straight chords. The lower bracket $h_n^-$ (orange) is the maximum of the trivial curve $e(\gamma)=(1-\gamma)_+$ and the supporting lines $\ell_{x,n}$ at the interior partition points (dashed gray). Convexity gives $h_n^-\le h\le h_n^+$, and all three curves agree at the partition points. The shaded region is the gap between the brackets. The example uses $h(\gamma)=(1-\gamma/2)_+^2$, $R=2$, and interior partition points $0.55$ and $1.30$.}
\label{fig:hockey-stick-brackets}
\end{figure}

\begin{proof}
Figure~\ref{fig:hockey-stick-brackets} illustrates the construction. For every $n\in\N$, choose partitions $0=\gamma_{0,n}<\cdots<\gamma_{m_n,n}=R$ with maximum interval length $\eta_n\to0$ as $n\to\infty$. Let $h_n^+$ be the linear interpolation of $h$ at the partition points, continued by zero after $R$. Convexity gives $h_n^+\ge h$, and its slopes lie in $[-1,0]$ and are nondecreasing, so $h_n^+$ is admissible.

For each interior partition point, choose a supporting line
\[
 \ell_{x,n}(\gamma)=h(\gamma_{x,n})+a_{x,n}(\gamma-\gamma_{x,n}),
 \qquad a_{x,n}\in[-1,0],\qquad\ell_{x,n}\le h.
\]
Such a slope can be chosen between the left and right derivatives of the convex function $h$. Define the lower bracket by
\be\label{eq:app-lower-bracket}
 h_n^-(\gamma)=\max\{e(\gamma),\ell_{1,n}(\gamma),\ldots,\ell_{m_n-1,n}(\gamma)\}.
\ee
Since $e\le h$, this is a piecewise-linear curve below $h$. As a maximum of affine functions with slopes in $[-1,0]$, it is convex, nonincreasing and $1$-Lipschitz. It equals $1$ at zero and vanishes for $\gamma\ge R$, so it too is admissible. Both brackets agree with $h$ at every partition point. Their $1$-Lipschitz bounds therefore give
\be\label{eq:app-bracket-error}
 0\le h_n^+-h_n^-,\qquad
 \|h_n^+-h_n^-\|_\infty\le2\eta_n\longrightarrow0.
\ee
Equation~\eqref{eq:app-polygonal-realization} supplies finite classical pairs $\X_n^\pm\in\mathsf L_R$ realizing the two brackets.

If $h=1-\gamma$ on $[0,\delta]$, include $\delta$ in every partition. The interpolation equals $h$ there, while $e\le h_n^-\le h$ forces the same equality for the lower bracket. The final assertion follows from~\eqref{eq:app-likelihood-curve}.
\end{proof}

In particular, if $\A=(\rho,\sigma)$ satisfies $\rho\le R\sigma$, then~\eqref{eq:app-admissible-brackets} gives $\X_n^+\succ_L\A\succ_L\X_n^-$. For full-rank $\rho,\sigma$, there is also a $\delta\in(0,1)$ with $\rho\ge\delta\sigma$, so the brackets can preserve both bounds. This is the approximation needed when the catalytic proof uses reverse relative entropy. The Lorenz-boundary interpretation and its figure are retained in Appendix~\ref{supp:curve-distance}.

\subsection{Proof of Theorem~\ref{unique-lorenz-extension}}\label{app:curve-completion}

Let $\D_{\rm cl}$ satisfy the hypotheses of Theorem~\ref{unique-lorenz-extension}. Classical pairs with the same curve Lorenz-majorize each other, so $\D_{\rm cl}$ depends only on that curve. For an admissible $h$, choose finite classical curves $h_{\X_n}\to h$ uniformly within one fixed upper-threshold sector, using Lemma~\ref{lem:bounded-classical-brackets}, and define
\be\label{eq:app-curve-completion}
 \G[h]\eqdef\lim_{n\to\infty}\D_{\rm cl}(\X_n).
\ee
Classical Lorenz continuity makes the values Cauchy and the limit independent of the approximating sequence: two sequences approaching $h$ have curve distance tending to zero. The same continuity modulus passes to the limit, so $\G$ is uniformly continuous on each fixed sector. Choosing a common larger threshold shows that the extensions agree on overlapping sectors.

If $h\ge k$, their classical brackets satisfy $h_{\X_n^+}\ge h\ge k\ge h_{\Y_n^-}$. Classical Lorenz monotonicity and passage to the limit give $\G[h]\ge\G[k]$. Thus $\G$ is pointwise monotone, and $\D(\A)=\G[h_{\A}]$ defines a QLD extension. Conversely, every QLD extension obeys
\be\label{eq:app-extension-squeeze}
 \D_{\rm cl}(\X_n^-)\le\D(\A)\le\D_{\rm cl}(\X_n^+)
 \qquad\text{when}\qquad h_{\X_n^-}\le h_{\A}\le h_{\X_n^+}\,.
\ee
Both bounds tend to $\G[h_{\A}]$, proving uniqueness and the equality of the envelopes~\eqref{eq:main-envelopes}. This proves Theorem~\ref{unique-lorenz-extension} and the representation~\eqref{eq:main-curve-functional}, including continuity on all admissible curves, without assuming quantum continuity.

\section{Classical Lorenz continuity}\label{supp:lorenz-continuity}

We use the uniform norm~\eqref{eq:main-uniform-norm} and Lorenz continuity from Definition~\ref{def:classical-lorenz-continuity}. This appendix supplies the continuity estimates used in the examples and the uniqueness proof. The equivalent Lorenz-boundary distance is treated separately in Appendix~\ref{supp:curve-distance}.

\subsection{Changing the first distribution}

We first record a simple Lipschitz estimate showing that, when the reference distribution is fixed, the uniform curve distance is controlled by the $\ell_1$ distance between the first distributions.
\bmyl\label{lipschitz}
Let $\p,\tp,\q\in\prob(n)$. Then:
\be\label{bo}
\|h_{(\p,\q)}-h_{(\tp,\q)}\|_\infty
\le
\|\p-\tp\|_1\, .
\ee
\emyl
\begin{proof}
For every $\gamma\ge0$,
\be
\left|
E_\gamma(\p\|\q)-E_\gamma(\tp\|\q)
\right|
=
\left|
\sum_x
\left[(p_x-\gamma q_x)_+-(\tilde p_x-\gamma q_x)_+\right]
\right|.
\ee
Using the elementary inequality
\be
|(a-c)_+-(b-c)_+|\le |a-b|,
\ee
we get
\be
\left|
E_\gamma(\p\|\q)-E_\gamma(\tp\|\q)
\right|
\le
\sum_x |p_x-\tilde p_x|
=
\|\p-\tp\|_1 .
\ee
Taking the supremum over $\gamma\ge0$ gives~\eqref{bo}.
\end{proof}

\subsection{R\'enyi divergences and their mixtures}

In Definition~\ref{def:classical-lorenz-continuity}, Lorenz continuity was defined as uniform continuity with respect to $\|\cdot\|_\infty$ on each bounded sector $\mathsf L_R$ defined in~\eqref{loc}. We now show that all classical R\'enyi divergences of finite order are Lorenz continuous.

\bmyl
\label{lem:renyi-lorenz-continuity}
For every $0<\alpha<\infty$, the classical R\'enyi divergence $D_\alpha$ is
Lorenz continuous. The endpoint divergences $D_0=D_{\min}$ and
$D_\infty=D_{\max}$ are not Lorenz continuous.
\emyl

\begin{proof}
Fix $R<\infty$ and let
$(\p,\q),(\tilde\p,\tilde\q)\in\mathsf L_R$. Set
\be
\delta
\coloneqq
\|h_{(\p,\q)}-h_{(\tp,\tq)}\|_\infty\,,\qquad
\Delta_\gamma
\eqdef
E_\gamma(\p\|\q)-E_\gamma(\tp\|\tq)\,.
\ee
Then $|\Delta_\gamma|\le\delta$ for all $\gamma\ge0$. Moreover,
$(\p,\q),(\tilde\p,\tilde\q)\in\mathsf L_R$ implies
\be
E_\gamma(\p\|\q)
=
h_{\tilde\X}(\gamma)
=
0
\qquad
\forall\,\gamma\ge R.
\ee
We first consider $\alpha\ne1$ and use the notation
$
Q_\alpha(\p\|\q)
\coloneqq
\rev{2^{(\alpha-1)D_\alpha(\p\|\q)}}
$.
The hockey-stick representation of $Q_\alpha$ is given by~\cite{LHC2025}
\be
Q_\alpha(\p\|\q)
=
1+\alpha(\alpha-1)
\int_1^\infty
\left[
\gamma^{\alpha-2}E_\gamma(\p\|\q)
+
\gamma^{-\alpha-1}E_\gamma(\q\|\p)
\right]\d\gamma .
\ee
Since $(\p,\q)\in\mathsf L_R$, the first term vanishes for $\gamma\ge R$.
For the second term we use
\be
E_\gamma(\q\|\p)
=
\gamma E_{\gamma^{-1}}(\p\|\q)+1-\gamma,
\qquad
\gamma>0.
\ee
Changing variables gives
\be
\int_1^\infty
\gamma^{-\alpha-1}E_\gamma(\q\|\p)\,\d\gamma
=
\int_0^1
\gamma^{\alpha-2}
\left[
E_\gamma(\p\|\q)-(1-\gamma)
\right]\d\gamma .
\ee
Thus
\be
Q_\alpha(\p\|\q)
=
1+\alpha(\alpha-1)
\Bigg[
\int_1^R
\gamma^{\alpha-2}
E_\gamma(\p\|\q)\,\d\gamma  
+
\int_0^1
\gamma^{\alpha-2}
\left(
E_\gamma(\p\|\q)-(1-\gamma)
\right)\d\gamma
\Bigg].
\ee
Since admissible curves satisfy $e\le h\le1$ (Section~\ref{sec:binary-curves}), both $E_\gamma(\p\|\q)$ and $E_\gamma(\tp\|\tq)$ lie in $[1-\gamma,1]$ for $0\le\gamma\le1$. Together with $|\Delta_\gamma|\le\delta$, this gives
\be
 |\Delta_\gamma|\le\min\{\delta,\gamma\},
 \qquad0\le\gamma\le1.
\ee
It follows that
\be
\Big|
Q_\alpha(\p\|\q)-Q_\alpha(\tilde\p\|\tilde\q)
\Big| \le
\alpha|\alpha-1|
\left[
\delta\int_1^R \gamma^{\alpha-2}\,\d\gamma
+
\int_0^1 \gamma^{\alpha-2}\min\{\delta,\gamma\}\,\d\gamma
\right]\,.
\ee
The first term is $O(\delta)$, while the second tends to zero as
$\delta\to0$ by dominated convergence, since
\be
\gamma^{\alpha-2}\min\{\delta,\gamma\}
\le
\gamma^{\alpha-1}.
\ee
Therefore $Q_\alpha$ is uniformly continuous on $\mathsf L_R$ with respect to
$\|\cdot\|_\infty$. Since
\be
\min\{1,R^{\alpha-1}\}
\le
Q_\alpha(\p\|\q)
\le
\max\{1,R^{\alpha-1}\},
\ee
the logarithm is uniformly continuous on the range of $Q_\alpha$. Hence
$D_\alpha=(\alpha-1)^{-1}\log Q_\alpha$ is Lorenz continuous.

For $\alpha=1$, the corresponding layer-cake formula is
\be
D_1(\p\|\q)
=
\rev{\frac{1}{\ln 2}}\int_0^1
\frac{E_\gamma(\p\|\q)-(1-\gamma)}{\gamma}\,\d\gamma
+
\rev{\frac{1}{\ln 2}}\int_1^R
\frac{E_\gamma(\p\|\q)}{\gamma}\,\d\gamma .
\ee
Consequently,
\be
\left|
D_1(\p\|\q)-D_1(\tilde\p\|\tilde\q)
\right|\le
\rev{\frac{1}{\ln 2}}\int_0^1
\frac{\min\{\delta,\gamma\}}{\gamma}\,\d\gamma
+
\rev{\frac{\delta}{\ln 2}}\int_1^R \frac{\d\gamma}{\gamma}\,.
\ee
For $0<\delta\le1$, this bound equals $\delta[1+\ln(R/\delta)]/\ln2$, which tends to zero. The same argument uses only admissibility, so the entropy integral in~\eqref{eq:main-entropy-integral} is uniformly continuous on every sector of admissible curves. Explicitly, for curves $h,k$ vanishing for $\gamma\ge R$ and $\delta=\|h-k\|_\infty\in(0,1]$,
\be\label{eq:app-entropy-curve-continuity}
 \frac1{\ln2}\left|\int_0^\infty\frac{h(\gamma)-k(\gamma)}\gamma\,\d\gamma\right|
 \le\frac{\delta}{\ln2}\bigl[1+\ln(R/\delta)\bigr].
\ee
This also proves the continuity of the functional $J$ used in the main proof.

It remains to show that the endpoint divergences are not Lorenz continuous.
For $D_{\max}$, let $\q_n=\tilde\p_n=\tilde\q_n=\u_n$, and set
\be
\p_n
=
\left(
\frac2n,
\frac{1-\frac2n}{n-1},
\ldots,
\frac{1-\frac2n}{n-1}
\right).
\ee
Then $(\p_n,\u_n),(\u_n,\u_n)\in\mathsf L_2$, and from Lemma~\ref{lipschitz}
\be
\|h_{(\p_n,\u_n)}-h_{(\u_n,\u_n)}\|_\infty
\le
\|\p_n-\u_n\|_1
\xrightarrow{n\to\infty}0\,.
\ee
However,
\be
D_{\max}(\p_n\|\u_n)=1\qquad\text{and}\qquad
D_{\max}(\u_n\|\u_n)=0\,.
\ee
Thus $D_{\max}$ is not Lorenz continuous.

For $D_{\min}$, let $n$ be even, let
$\q_n=\tilde\q_n=\u_n$, and let $\{\eps_n\}_{n\in\N}$ be a sequence of positive numbers with a zero limit. Define
\be
\p_n
=
\Big(
\underbrace{\frac2n,\ldots,\frac2n}_{n/2},
\underbrace{0,\ldots,0}_{n/2}
\Big),
\ee
and
\be
\tilde\p_n
=
\Big(
\underbrace{\frac{2-\eps_n}{n},\ldots,
\frac{2-\eps_n}{n}}_{n/2},
\underbrace{\frac{\eps_n}{n},\ldots,
\frac{\eps_n}{n}}_{n/2}
\Big).
\ee
Then $(\p_n,\u_n),(\tp_n,\u_n)\in\mathsf L_2$, and
\be
\|h_{(\p_n,\u_n)}-h_{(\tp_n,\u_n)}\|_\infty
\le
\|\p_n-\tilde\p_n\|_1
=
\eps_n
\longrightarrow0.
\ee
On the other hand,
\be
D_{\min}(\p_n\|\u_n)=\log (2)
\qquad\quad\text{and}\qquad
D_{\min}(\tilde\p_n\|\u_n)=0.
\ee
Thus $D_{\min}$ is not Lorenz continuous.
\end{proof}

\noindent\textbf{Remarks.}
\ben
\item For any fixed probability measure $\pi$ on $(0,\infty)$, let $\D_\pi=\int D_\alpha\,\d\pi(\alpha)$. On $\mathsf L_R$, every order satisfies $0\le D_\alpha\le\log R$. If two sequences of pairs have curve distances tending to zero in that sector, Lemma~\ref{lem:renyi-lorenz-continuity} makes their divergence differences tend to zero for each fixed $\alpha$. The bound $\log R$ permits dominated convergence over $\pi$, so their $\D_\pi$ differences also tend to zero. This sequential criterion proves uniform Lorenz continuity of every fixed probability mixture. Zero normalization and classical-catalytic monotonicity pass directly through the same positive integral. The associated entropy is the corresponding probability mixture of R\'enyi entropies, so expansibility on uniform distributions is also preserved.
\item Uniform convergence of the curves alone does not control arbitrarily large likelihood ratios on events of vanishing reference probability. Let $\u_n$ be uniform on $n$ outcomes, put $a_n=1/\log n$, and set
\be
 \p_n=\left(a_n,\frac{1-a_n}{n-1},\ldots,\frac{1-a_n}{n-1}\right).
\ee
Then $\|h_{(\p_n,\u_n)}-e\|_\infty\le\|\p_n-\u_n\|_1\to0$, while
\be
 D(\p_n\|\u_n)=a_n\log(na_n)+(1-a_n)\log\frac{n(1-a_n)}{n-1}\longrightarrow1.
\ee
Thus even relative entropy is not uniformly continuous across unbounded sectors. The definition asks only for stability within each fixed finite upper threshold.
\een

\subsection{Classical \texorpdfstring{$f$}{f}-divergences}

Let $f:[0,\infty)\to\mathbb R$ be finite and convex, and continuous at zero. For a classical pair $\X=(\p,\q)\in\mathsf L_R$, its $f$-divergence is
\be\label{eq:app-classical-f-divergence}
 D_f(\X)=\sum_{x:q_x>0}q_x f\!\left(\frac{p_x}{q_x}\right).
\ee
All likelihood ratios belong to $[0,R]$. We will prove uniform continuity on this sector by approximating $f$ with piecewise-linear convex functions, whose divergences depend on finitely many values of $h_{\X}$.

Choose partitions $0=t_{0,n}<\cdots<t_{m_n,n}=R$ containing $1$, with maximum interval length tending to zero, and let $f_n$ interpolate $f$ linearly between successive partition points. Convexity makes the successive slopes nondecreasing, so $f_n$ is convex. Since $f$ is continuous on $[0,R]$, these interpolants satisfy
\be\label{eq:app-f-uniform-approximation}
 \varepsilon_n\eqdef\sup_{0\le t\le R}|f(t)-f_n(t)|\longrightarrow0.
\ee
Indeed, on each partition interval the interpolation error is bounded by the maximum variation of $f$ on that interval. Including $1$ also preserves the usual normalization $f_n(1)=f(1)=0$ when $f(1)=0$.

Let $a_n=f(0)$, let $b_n$ be the slope on the first interval, and let $c_{j,n}\ge0$ be the increase in slope at the interior point $t_{j,n}$. Then
\be\label{eq:app-f-polygonal-representation}
 f_n(t)=a_n+b_nt+\sum_{j=1}^{m_n-1}c_{j,n}(t-t_{j,n})_+,
 \qquad0\le t\le R.
\ee
Each hockey stick contributes precisely the required change in slope. Substituting into~\eqref{eq:app-classical-f-divergence}, using $\sum_xp_x=\sum_xq_x=1$ and~\eqref{eq:app-likelihood-curve}, gives
\be\label{eq:app-f-polygonal-divergence}
 D_{f_n}(\X)=a_n+b_n+\sum_{j=1}^{m_n-1}c_{j,n}h_{\X}(t_{j,n}).
\ee
Consequently, for any $\X,\Y\in\mathsf L_R$,
\be\label{eq:app-f-polygonal-lipschitz}
 |D_{f_n}(\X)-D_{f_n}(\Y)|
 \le L_n\|h_{\X}-h_{\Y}\|_\infty,
 \qquad L_n\eqdef\sum_{j=1}^{m_n-1}c_{j,n}<\infty.
\ee
Thus each approximating divergence is Lipschitz in the curve norm.

The approximation error is also uniform over all pairs in the sector:
\be\label{eq:app-f-divergence-error}
 |D_f(\X)-D_{f_n}(\X)|
 \le\sum_{x:q_x>0}q_x\left|f\!\left(\frac{p_x}{q_x}\right)-f_n\!\left(\frac{p_x}{q_x}\right)\right|
 \le\varepsilon_n\sum_xq_x=\varepsilon_n.
\ee
Combining~\eqref{eq:app-f-polygonal-lipschitz} and~\eqref{eq:app-f-divergence-error} yields
\be\label{eq:app-f-continuity-bound}
 |D_f(\X)-D_f(\Y)|
 \le2\varepsilon_n+L_n\|h_{\X}-h_{\Y}\|_\infty.
\ee
Given any tolerance $\eta>0$, first choose $n$ with $2\varepsilon_n<\eta/2$, then require $\|h_{\X}-h_{\Y}\|_\infty<\eta/(2\max\{1,L_n\})$. The divergence difference is then smaller than $\eta$. Both choices depend only on $f$, $R$ and $\eta$, not on the alphabet sizes. This proves Lorenz continuity of $D_f$, as required for its unique HT extension in Section~\ref{sec:extension}.

\section{Consequences of expansibility on uniform distributions}\label{app:expansibility}

This appendix translates the entropy condition~\eqref{eq:main-uniform-expansibility} into the values on elementary classical pairs that will be used in the uniqueness proof. It also shows that expansibility does not imply classical additivity, even for a normalized Lorenz-continuous divergence.

\subsection{Elementary pairs and logarithmic calibration}\label{app:expansibility-calibration}

\bmyl[Elementary consequence of expansibility]\label{lem:expansibility-calibration}
Let $\D$ be a classical Lorenz-monotone divergence with $\D(1\|1)=0$, and let $\H$ be its associated entropy~\eqref{eq:main-associated-entropy}. Expansibility on uniform distributions is equivalent to
\be\label{eq:cat-calibration}
 \D\bigl((1,0)\|(t,1-t)\bigr)=-\log (t),
 \qquad0<t\le1.
\ee
It suffices to impose expansibility for embeddings into dimensions $d=2^n$. No continuity assumption is needed for this equivalence.
\emyl

\begin{proof}
DPI under discarding and state preparation gives $\D(\mathbf u_k\|\mathbf u_k)=0$, hence $\H(\mathbf u_k)=\log (k)$. For $1\le k\le d$, set
\be\label{eq:app-uniform-pair}
 \p_{k,d}\eqdef\mathbf u_k\oplus\mathbf0_{d-k},
 \qquad\Y_{k,d}\eqdef(\p_{k,d},\mathbf u_d).
\ee
Expansibility gives
\ba\label{eq:app-uniform-value}
 \D(\Y_{k,d})
 &=\log(d)-\H(\mathbf u_k\oplus\mathbf0_{d-k})\\
 &=\log(d)-\H(\mathbf u_k)=\log(d/k).
\ea
Introduce the elementary binary pair
\be\label{eq:app-binary-pair}
 \X_t\eqdef\bigl((1,0),(t,1-t)\bigr),\qquad0<t\le1.
\ee
The two curves satisfy
\be\label{eq:app-uniform-binary-equivalence}
 h_{\Y_{k,d}}(\gamma)
 =k\left(\frac1k-\frac\gamma d\right)_+
 =\left(1-\frac{k}{d}\gamma\right)_+
 =h_{\X_{k/d}}(\gamma).
\ee
Thus $\Y_{k,d}$ and $\X_{k/d}$ Lorenz-majorize each other and have the same divergence. Operationally, grouping the first $k$ coordinates into one outcome gives $\X_{k/d}$; uniformly splitting each of the two groups reconstructs $\Y_{k,d}$ when $k<d$. For $k=d$, both hypotheses are identical. Equation~\eqref{eq:app-uniform-value} therefore proves~\eqref{eq:cat-calibration} for rational $t=k/d$.

In fact, dimensions $d=2^n$ already give all dyadic probabilities $k/2^n$, which are dense in $(0,1)$. For an arbitrary $0<t<1$, choose dyadic $s_n<t<r_n$ with $s_n,r_n\to t$. Since $h_{\X_s}(\gamma)=(1-s\gamma)_+$ decreases with $s$, Lorenz monotonicity gives
\be\label{eq:app-calibration-squeeze}
 -\log(r_n)=\D(\X_{r_n})
 \le\D(\X_t)
 \le\D(\X_{s_n})=-\log(s_n)\,.
\ee
The two bounds converge to $-\log t$. At $t=1$, both hypotheses coincide and the divergence is zero. This establishes~\eqref{eq:cat-calibration} without invoking continuity of $\D$.

Conversely,~\eqref{eq:cat-calibration} and~\eqref{eq:app-uniform-binary-equivalence} give $\D(\Y_{k,d})=\log(d/k)$. Substituting into~\eqref{eq:main-associated-entropy} yields
\[
 \H(\mathbf u_k\oplus\mathbf0_{d-k})
 =\log(d)-\log(d/k)=\log k=\H(\mathbf u_k),
\]
which is expansibility on uniform distributions.
\end{proof}

Equation~\eqref{eq:cat-calibration} is often interpreted as logarithmic calibration: an event certain under the first hypothesis and of probability $t$ under the second provides $-\log (t)$ bits of evidence. Here this formula is a consequence of the entropy assumption in the main text. In particular, $k=1,d=2$ gives $\D((1,0)\|(1/2,1/2))=1$; a separate one-bit normalization is unnecessary. The base of the logarithm in~\eqref{eq:main-associated-entropy} fixes the units.

\subsection{Relation to additivity}\label{app:expansibility-additivity}

\bmyl[Additivity implies expansibility]\label{lem:app-additive-expansibility}
A real-valued classical Lorenz divergence that is additive on independent pairs and satisfies $\D((1,0)\|(1/2,1/2))=1$ obeys $\D(1\|1)=0$ and expansibility on uniform distributions.
\emyl

\begin{proof}
Additivity on the trivial pair gives $\D(1\|1)=2\D(1\|1)$, hence $\D(1\|1)=0$. For the pairs in~\eqref{eq:app-binary-pair}, direct calculation gives $h_{\X_s\otimes\X_t}=h_{\X_{st}}$. Therefore $f(t)\eqdef\D(\X_t)$ satisfies
\be\label{eq:app-calibration-counting}
 f(st)=f(s)+f(t),\qquad f(1/2)=1.
\ee
The function $a(r)\eqdef f(2^{-r})$, $r\ge0$, is nondecreasing by Lorenz monotonicity and satisfies $a(r+s)=a(r)+a(s)$ and $a(1)=1$. Additivity first gives $a(r)=r$ for nonnegative rational $r$; monotonicity and rational approximation give the same equality for every $r\ge0$. Thus $f(t)=-\log (t)$, and Lemma~\ref{lem:expansibility-calibration} gives expansibility on uniform distributions.
\end{proof}

\subsection{Expansibility does not imply additivity}\label{app:expansibility-nonadditivity}

An explicit classical counterexample shows that the converse fails. Let $D_1$ and $D_2$ denote the classical R\'enyi divergences of orders one and two, and define
\be\label{eq:app-nonadditive-divergence}
 \D_\star(\p\|\q)\eqdef
 \log\!\left(\frac{2^{D_1(\p\|\q)}+2^{D_2(\p\|\q)}}2\right).
\ee
Both R\'enyi divergences obey classical DPI and tensor-product additivity~\cite{VanErvenHarremoes2014}. Since~\eqref{eq:app-nonadditive-divergence} is increasing in both arguments, $\D_\star$ obeys DPI and hence classical Lorenz monotonicity. It vanishes on identical distributions and is Lorenz continuous on each bounded sector: both components have this property by Lemma~\ref{lem:renyi-lorenz-continuity}, and the displayed function is uniformly continuous on their bounded ranges. Moreover, $D_1(\X_t)=D_2(\X_t)=-\log (t)$, so $\D_\star$ satisfies~\eqref{eq:cat-calibration} and therefore expansibility on uniform distributions.

The associated entropy even satisfies expansibility for every distribution. Indeed,
\be\label{eq:app-nonadditive-entropy}
 \H_\star(\p)
 =-\log\!\left(\frac{2^{-H_1(\p)}+2^{-H_2(\p)}}2\right),
\ee
where $H_1(\p)=-\sum_xp_x\log (p_x)$ and $H_2(\p)=-\log\sum_xp_x^2$. These two entropies are unchanged by appending zeros, so $\H_\star$ is also unchanged; it attains its maximum $\log(d)$ at $\mathbf u_d$.

Nevertheless, write $a=2^{D_1(\p\|\q)}$ and $b=2^{D_2(\p\|\q)}$. Additivity of the two component divergences gives
\ba\label{eq:app-nonadditivity-gap}
 &\D_\star(\p^{\otimes2}\|\q^{\otimes2})-2\D_\star(\p\|\q)\\
 &\qquad=\log\!\left(\frac{2(a^2+b^2)}{(a+b)^2}\right)
 =\log\!\left(1+\frac{(a-b)^2}{(a+b)^2}\right)>0
\ea
whenever $D_1(\p\|\q)\ne D_2(\p\|\q)$.  Thus even full expansibility of the associated entropy, together with zero normalization and Lorenz continuity, does not imply classical additivity. In Theorem~\ref{thm:cat-uniqueness}, additivity follows only after the quantum catalytic principle has selected Umegaki relative entropy.

\clearpage

\section{Proof of main result}\label{app:main-proof}

We prove Theorem~\ref{thm:cat-uniqueness}. The argument first reduces uniqueness to equality on joint entropy level sets, then derives the replacements permitted by catalytic monotonicity, and finally constructs the required changes using full-rank qubit pairs.

We use admissible curves, the trivial curve $e$, and the uniform norm as introduced in Section~\ref{sec:binary-curves}; see Definition~\ref{def:cat-admissible} and~\eqref{eq:main-uniform-norm}. For two admissible curves $h_0$ and $h_1$ we write $h_0\ge h_1$ if  $h_0(\gamma)\ge h_1(\gamma)$ for every $\gamma\ge0$. The classical representation, brackets and continuous extension have been established in Appendix~\ref{supp:classical-brackets}, and the required continuity estimates in Appendix~\ref{supp:lorenz-continuity}.

For curve differences, $\|f\|_\infty=\sup_{\gamma\ge0}|f(\gamma)|$ for any bounded real function $f$ on $[0,\infty)$. For a scalar function $f$, its \emph{support} is
\[
 \supp f\eqdef\overline{\{\gamma\ge0:f(\gamma)\ne0\}},
\]
where the bar denotes closure: the set includes all limit points of points where $f$ is nonzero. Thus $\supp f\subseteq I$ means that $f$ vanishes outside the closed set $I$. A compact interval $I\subset(0,\infty)$ will mean $I=[a,b]$ with $0<a<b<\infty$; its supremum $\sup I$ is its least upper bound, namely the right endpoint $b$.

Catalytic Lorenz monotonicity implies ordinary Lorenz monotonicity by choosing a catalyst $\C=(\tau,\tau)$, which leaves every hockey-stick curve unchanged. Theorem~\ref{unique-lorenz-extension} and its completion argument in Appendix~\ref{supp:classical-brackets} therefore give
$\D(\A)=\G[h_{\A}]$, where $\G$ is defined on the admissible curves and is pointwise monotone and uniformly continuous with respect to $\|\cdot\|_\infty$ on each fixed upper-threshold sector. No quantum continuity is assumed.
Here \emph{pointwise monotonicity} means that, for any admissible curves $h_0,h_1$,
\be
 h_0\ge h_1
 \quad\Longrightarrow\quad\G[h_0]\ge\G[h_1].
\ee
Pointwise monotonicity of $\G$ is Lorenz monotonicity of $\D$ expressed in terms of hockey-stick curves, extended to all admissible curves by classical approximation and continuity. This property compares the values of $\G$ on two curves; the nonincreasing condition in admissibility describes each individual curve as $\gamma$ varies.

A fixed upper-threshold sector consists of curves that all vanish for $\gamma\ge R$, with one common $R$. When a positive lower likelihood-ratio bound is also needed, we use the initial-segment condition in Lemma~\ref{lem:bounded-classical-brackets}; the lower bound may depend on the curve.

The entropy integral formula gives the affine coordinates
\begin{equation}\label{eq:cat-entropies}
 J[h]=\frac1{\ln2}\int_0^\infty\frac{h(\gamma)-e(\gamma)}\gamma\,\d\gamma,
 \qquad
 K[h]=\frac1{\ln2}\int_0^\infty\frac{h(\gamma)-e(\gamma)}{\gamma^2}\,\d\gamma.
\end{equation}
Since $h\ge e$, both integrands are nonnegative; in particular, $J[h]\ge0$. Equation~\eqref{eq:app-entropy-curve-continuity} proves that $J$ is uniformly continuous on every fixed upper-threshold sector.
Here $J[h_{\A}]=D(\rho\|\sigma)$ and, for equal supports, $K[h_{\A}]=D(\sigma\|\rho)$. Both are additive on products when finite; this is a property of the known entropy, not of $\D$. Logarithms are base two; $\ln$ is natural.
If $h(\gamma)=1-\gamma$ near zero, both integrands in \eqref{eq:cat-entropies} vanish there and beyond a finite upper threshold, so $J[h]$ and $K[h]$ are finite.

It suffices to establish the representation $\G[h]=g(J[h])$. Lemma~\ref{lem:expansibility-calibration} has already derived~\eqref{eq:cat-calibration} from zero normalization and expansibility on uniform distributions. For every $r\ge0$, the pair $\X_{2^{-r}}$ in~\eqref{eq:app-binary-pair} has curve $\gamma\mapsto(1-2^{-r}\gamma)_+$ and satisfies
\be\label{eq:app-expansibility-fixes-g}
 J[h_{\X_{2^{-r}}}]=r,
 \qquad g(r)=\G[h_{\X_{2^{-r}}}]=\D(\X_{2^{-r}})=r.
\ee
Hence $g$ is the identity, $\G[h]=J[h]$, and $\D=D$. The remaining proof establishes the representation with $g$ continuous and nondecreasing, without using normalization or expansibility.

\subsection{Reduce to two fixed entropy values}\label{app:cat-reduction}

Our goal is to show that $\G[h]$ depends only on $J[h]$: for any two admissible hockey-stick curves $h_0,h_1$, we need to prove that $J[h_0]=J[h_1]$ implies $\G[h_0]=\G[h_1]$. The following lemma shows that it suffices to prove this when the curves also have the same $K$ value and agree with $1-\gamma$ near zero. It also establishes the continuity and monotonicity of the resulting function $g$.

\bmyl[Reduction to two fixed entropy values]\label{lem:cat-reduction}
Let $\G$ be pointwise monotone and uniformly continuous on each fixed upper-threshold sector of admissible hockey-stick curves. Suppose that
\begin{equation}\label{eq:cat-levelsets}
 J[h_0]=J[h_1],\quad K[h_0]=K[h_1]
 \quad\Longrightarrow\quad\G[h_0]=\G[h_1]\,,
\end{equation}
for all admissible curves equal to $1-\gamma$ near zero (i.e., satisfying $h_0(\gamma)=h_1(\gamma)=1-\gamma$ on $[0,\delta]$ for some $\delta\in(0,1)$). Then there is a continuous nondecreasing function $g:[0,\infty)\to\mathbb R$ such that $\G[h]=g(J[h])$ for every admissible curve $h$.
\emyl

\begin{proof}
First consider curves $h_0,h_1$ equal to $1-\gamma$ near zero, so both $J$ and $K$ are finite. Suppose $J[h_0]=J[h_1]=\lambda>0$. If their $K$ values differ, relabel the curves so that $K[h_0]=v<K[h_1]=w$; equal $K$ is already covered by the hypothesis~\eqref{eq:cat-levelsets}. Fix $r>\max\{1,\lambda\}$ and let $\X=(\p,\q)$, where
\be
 h_{\X}(\gamma)=E_\gamma(\p\|\q),\qquad
 \p=(\eps,1-\eps),\quad\q=(1-2^{-r},2^{-r}),\qquad0<\eps<1.
\ee
Then
\ba\label{eq:cat-binary-entropy-limits}
 \mu\eqdef J[h_{\X}]=D(\p\|\q)&\xrightarrow{\eps\to0^+}r>\lambda,\\
 K[h_{\X}]=D(\q\|\p)&\xrightarrow{\eps\to0^+}\infty.
\ea
Since $p_x/q_x\le2^r$ for $x=1,2$, the curve $h_{\X}$ vanishes for $\gamma\ge2^r$. For sufficiently small $\eps$, we have $\mu>\lambda$. Mixing with the trivial curve $e$ therefore gives
\be\label{eq:cat-dilution}
 f(\gamma)=\frac{\lambda}{\mu}h_{\X}(\gamma)
       +\left(1-\frac{\lambda}{\mu}\right)e(\gamma)\,.
\ee
By affinity of $J$ and $K$, both of which vanish on the trivial curve $e$,
\ba
 J[f]&=\frac{\lambda}{\mu}\mu+\left(1-\frac{\lambda}{\mu}\right)0=\lambda,\\
 K[f]&=\frac{\lambda}{\mu}K[h_{\X}]+\left(1-\frac{\lambda}{\mu}\right)0
 =\frac{\lambda}{\mu}K[h_{\X}]\xrightarrow{\eps\to0^+}\infty,
\ea
where the divergence follows from \eqref{eq:cat-binary-entropy-limits} and $\lambda/\mu\to\lambda/r>0$.
Here $\X,\p,\mu,f$, and the weight below depend on $\eps$, while $r$ and $\q$ remain fixed.

Choose $\eps$ small enough that $K[f]>w$, and set
\be
 \delta=\frac{w-v}{K[f]-v}\in(0,1),\qquad
 \widetilde h_\eps=(1-\delta)h_0+\delta f\,.
\ee
Using again the affinity of $J$ and $K$, we obtain
\ba
 J[\widetilde h_\eps]&=(1-\delta)\lambda+\delta\lambda=\lambda,\\
 K[\widetilde h_\eps]&=(1-\delta)v+\delta K[f]=w.
\ea
Consequently, $\widetilde h_\eps$ and $h_1$ have the same $J$ and $K$ values. Since both curves also agree with $1-\gamma$ near zero, the hypothesis~\eqref{eq:cat-levelsets} gives
$\G[\widetilde h_\eps]=\G[h_1]$.
Moreover, $\delta\to0$, and since both $f$ and $h_0$ take values in $[0,1]$,
\be
 \|\widetilde h_\eps-h_0\|_\infty
 =\delta\|f-h_0\|_\infty\le\delta\xrightarrow{\eps\to0^+}0.
\ee
For each such $\eps$, the curve $f$ vanishes for $\gamma\ge2^r$, so the mixtures $\widetilde h_\eps$ remain in one fixed bounded sector. Lorenz continuity now yields $\G[h_0]=\G[h_1]$. If $\lambda=0$, nonnegativity of $h(\gamma)-e(\gamma)$ and continuity of $h$ imply that $J[h]=0$ only for $h=e$, so equality is immediate.

We have proved that $\G$ is constant on each $J$-level set among curves equal to $1-\gamma$ near zero. Every $\lambda\ge0$ is attained by a curve in this class. Indeed, for any prescribed $\lambda>0$, choose $r>\max\{1,\lambda\}$ and then $\eps>0$ small enough that $\mu>\lambda$, which is possible because $\mu\to r$ in \eqref{eq:cat-binary-entropy-limits}. The curve $f$ in \eqref{eq:cat-dilution} is then a convex mixture of curves equal to $1-\gamma$ near zero and satisfies $J[f]=(\lambda/\mu)\mu=\lambda$. The trivial curve $e$ attains $\lambda=0$.

We may therefore define $g(\lambda)=\G[h]$ for every $\lambda\ge0$ by choosing any curve $h$ in this class with $J[h]=\lambda$. Such a curve exists by the construction in \eqref{eq:cat-dilution}, and constancy on each $J$-level set makes the value independent of the choice.

To prove continuity and monotonicity of $g$, now fix $r$ and $\eps$, so that $\X$ and $\mu$ are fixed, and let $\lambda$ vary over $[0,\mu]$ in \eqref{eq:cat-dilution}. Write $f_\lambda$ for this curve to make its dependence on $\lambda$ explicit. Then $J[f_\lambda]=\lambda$, and
\[
 \|f_\lambda-f_{\lambda'}\|_\infty
 =\frac{|\lambda-\lambda'|}{\mu}\,\|h_{\X}-e\|_\infty
 \le\frac{|\lambda-\lambda'|}{\mu}.
\]
Thus $f_{\lambda'}\to f_\lambda$ uniformly as $\lambda'\to\lambda$. All these curves vanish above $2^r$, so Lorenz continuity of $\G$ makes $g(\lambda)=\G[f_\lambda]$ continuous on $[0,\mu]$. The family also increases pointwise with $\lambda$, because $h_{\X}(\gamma)\ge e(\gamma)$; monotonicity of $\G$ therefore makes $g$ nondecreasing. Since $r$, and hence the available values of $\mu$, can be arbitrarily large, both conclusions hold on all of $[0,\infty)$.

Finally, let $h$ be any admissible curve. We approximate it by curves $h_n$ equal to $1-\gamma$ near zero, all in a common bounded sector. To construct them, use Lemma~\ref{lem:bounded-classical-brackets} to choose finite classical pairs $\X_n=(\p_n,\q_n)$ whose curves satisfy $h_{\X_n}\to h$ (as $n\to\infty$) uniformly within such a sector, choose $\delta_n\in(0,1)$ tending to zero, and set
\be
 \Y_n\eqdef\bigl((1-\delta_n)\p_n+\delta_n\q_n,\q_n\bigr),\qquad h_n\eqdef h_{\Y_n}.
\ee
The new likelihood ratios satisfy
\be
 \frac{(1-\delta_n)p_{n,x}+\delta_n q_{n,x}}{q_{n,x}}
 =(1-\delta_n)\frac{p_{n,x}}{q_{n,x}}+\delta_n\ge\delta_n\qquad(q_{n,x}>0),
\ee
so $h_n(\gamma)=1-\gamma$ on $[0,\delta_n]$. The common upper threshold is preserved, and Lemma~\ref{lipschitz} gives $\|h_n-h_{\X_n}\|_\infty\le\delta_n\|\q_n-\p_n\|_1\le2\delta_n$, hence $\|h_n-h\|_\infty\to0$. The representation already proved applies to each $h_n$. Continuity of $\G$, $J$, and $g$ now gives
\be
 \G[h]=\lim_{n\to\infty}\G[h_n]
 =\lim_{n\to\infty}g(J[h_n])=g(J[h]).
\ee
Thus the representation holds for every admissible curve, even when the limit no longer equals $1-\gamma$ near zero.

\end{proof}

\subsection{Classical products of hockey-stick curves}\label{app:classical-product}

By Lemma~\ref{lem:cat-reduction}, it remains to prove that curves with the same $J$ and $K$ values have the same $\G$ value, provided they equal $1-\gamma$ near zero. To complete this reduction, we must exploit catalytic monotonicity, which compares tensor products of pairs. We first introduce a product of admissible curves that exactly describes tensoring whenever one factor is classical. This section establishes its properties. The next section uses classical approximations and catalytic monotonicity to relate this product to tensor products of quantum pairs.

\bmyd[Classical product]\label{def:cat-classical-product}
For an admissible curve $h$ and the curve $k=h_{\X}$ of a finite classical pair $\X=(\p,\q)$, define their \emph{classical product} by
\begin{equation}\label{eq:cat-classical-product}
 (h\odot k)(\gamma)\eqdef\sum_{x:p_x>0}p_x\,
 h\!\left(\frac{\gamma q_x}{p_x}\right).
\end{equation}
For arbitrary admissible curves, we define $h\odot k$ by uniform classical approximation; the existence, uniqueness and properties of this extension are justified below.
\emyd

\paragraph{Properties of the classical product.}
For any quantum pair $\A$ and classical pair $\X=(\p,\q)$,
\begin{equation}\label{eq:cat-mixed-product}
 h_{\A\otimes \X}(\gamma)
 =\sum_{x:p_x>0}p_xh_{\A}\!\left(\frac{\gamma q_x}{p_x}\right)
 =(h_{\A}\odot h_{\X})(\gamma),
\end{equation}
because the classical register makes the operator block diagonal, so the positive part and trace can be evaluated separately in each block.
Blocks with $p_x=0$ contribute zero.

For finite classical pairs $\X,\Y$, equation~\eqref{eq:cat-mixed-product} gives
\be
 h_{\X}\odot h_{\Y}=h_{\X\otimes \Y}.
\ee
Permuting or regrouping the tensor factors therefore proves commutativity and associativity on finite classical curves. The approximation argument below extends these properties to all admissible curves, including those arising from quantum pairs.

These observations also show that $\odot$ is well defined on finite
classical curves: the defining formula depends on the first pair only
through its curve, and commutativity gives the same conclusion for
the second pair. Moreover, for fixed classical $k$,
\eqref{eq:cat-classical-product} preserves convex mixtures in $h$.
Its nonnegative weights also preserve pointwise inequalities:
$h\ge\widetilde h$ implies $h\odot k\ge\widetilde h\odot k$.
By commutativity, both properties hold in the second argument as well.
Thus $\odot$ is affine and order preserving in each argument on
finite classical curves.

For fixed classical $k$, the weights sum to one, so \eqref{eq:cat-classical-product} also gives
\be\label{21}
 \|h\odot k-\widetilde h\odot k\|_\infty
 \le\sum_{x:p_x>0}p_x\|h-\widetilde h\|_\infty
 =\|h-\widetilde h\|_\infty\,.
\ee
Applying this bound in each argument and using the triangle inequality yields
\ba\label{22}
 \|h\odot k-\widetilde h\odot\widetilde k\|_\infty
 &\le\|h\odot k-\widetilde h\odot k\|_\infty
       +\|\widetilde h\odot k-\widetilde h\odot\widetilde k\|_\infty\\
 &\le\|h-\widetilde h\|_\infty+\|k-\widetilde k\|_\infty\,.
 \ea
The upper thresholds are controlled as well: if $h$ vanishes for $\gamma\ge R$ and $k$ for $\gamma\ge S$, then $p_x/q_x\le S$. For $\gamma\ge RS$, every argument $\gamma q_x/p_x$ in \eqref{eq:cat-classical-product} is at least $R$, so $h\odot k$ vanishes there.

For general admissible $h,k$, choose finite classical approximations $h_n\to h$ and $k_n\to k$ uniformly with common upper thresholds. The estimate~\eqref{22} makes $h_n\odot k_n$ uniformly Cauchy, with a limit independent of the approximations; this defines $h\odot k$. The limit is admissible, since the defining curve properties and the common upper threshold are preserved. Commutativity, associativity, affinity and the estimate pass to the limit. When $k$ is classical, this extension agrees with \eqref{eq:cat-classical-product}, by the one-argument estimate~\eqref{21}. The formula in~\eqref{eq:cat-classical-product} gives order preservation for arbitrary $h\ge\widetilde h$ and classical $k$; approximating $k$ proves it for all admissible curves.

\subsection{Catalytic replacement}\label{app:catalytic-replacement}

We now use the additional constraint supplied by catalytic monotonicity. We will show that replacing $h_{\A}\odot h_{\C}$ by $h_{\A\otimes \C}$, even inside a common convex mixture, leaves $\G$ unchanged. Section~\ref{app:replacement-combinations} combines these replacements to prove constancy on the joint $J,K$ level sets required by Lemma~\ref{lem:cat-reduction}.

Let $\A_0=(\rho_0,\sigma_0)$ and $\A_1=(\rho_1,\sigma_1)$ be arbitrary state pairs, and let $\C$ be a catalyst pair. Under the representation $\D(\A_0)=\G[h_{\A_0}]$, catalytic monotonicity says precisely that
\begin{equation}\label{eq:cat-functional-monotonicity}
 h_{\A_0\otimes \C}(\gamma)\ge h_{\A_1\otimes \C}(\gamma)\ \text{for all }\gamma\ge0
 \quad\Longrightarrow\quad\G[h_{\A_0}]\ge\G[h_{\A_1}].
\end{equation}
We use no general characterization of catalytic Lorenz majorization: we construct particular comparisons with an explicit catalyst.

For quantum pairs $\A,\C$, the actual product curve $h_{\A\otimes \C}$ need not equal $h_{\A}\odot h_{\C}$. We will see an explicit full-rank qubit example in Section~\ref{app:qubit-calculation}: \eqref{eq:cat-exact-area} shows that these two curves have different integrals. Classical bracketing does not force equality, because Lorenz dominance need not be preserved when both pairs are tensored with the same quantum pair. Indeed, if classical curves $h_{\Z_n}$ converge uniformly to $h_{\A}$, then $h_{\Z_n\otimes \C}=h_{\Z_n}\odot h_{\C}$ converges to $h_{\A}\odot h_{\C}$; this gives no conclusion about $h_{\A\otimes \C}$. Thus the output curves in \eqref{eq:cat-functional-monotonicity} cannot in general be replaced by classical products of their input curves.

To formulate the common-mixture statement, we also need to realize convex mixtures of curves by state pairs. A direct sum with the same block probabilities under both hypotheses does exactly this: for $\A_0=(\rho_0,\sigma_0)$, $\A_1=(\rho_1,\sigma_1)$ and $0\le t\le1$,
\ba\label{eq:cat-mixture}
 \B_t&\eqdef\big((1-t)\rho_0\oplus t\rho_1,\,(1-t)\sigma_0\oplus t\sigma_1\big),\\
 h_{\B_t}(\gamma)&=(1-t)h_{\A_0}(\gamma)+t h_{\A_1}(\gamma).
\ea
The second identity follows because the positive part and trace act separately on each block.

\bmyl\label{lem:cat-replacement}
For any two state pairs $\A,\C$,
\begin{equation}\label{eq:cat-product-invariance}
\G[h_{\A\otimes \C}]=\G[h_{\A}\odot h_{\C}].
\end{equation}
More generally, for every admissible curve $k$ and $0\le t\le1$,
\begin{equation}\label{eq:cat-replacement}
 \G[(1-t)k+t(h_{\A}\odot h_{\C})]
 =\G[(1-t)k+t h_{\A\otimes \C}].
\end{equation}
\emyl

The mixture identity is essential because no affinity of $\G$ is assumed: it allows us to replace $h_{\A}\odot h_{\C}$ by $h_{\A\otimes \C}$ within any convex mixture while leaving its other components unchanged. Section~\ref{app:replacement-combinations} uses successive replacements of this kind to combine their effects.

\begin{proof}
We use $\C$ as the catalyst and approximate $\C$ and $\A\otimes \C$ by finite classical brackets. The key identity, valid for every finite classical pair $\X$, is
\begin{equation}\label{eq:cat-exchange}
 h_{(\A\otimes \X)\otimes \C}
 =h_{(\A\otimes \C)\otimes \X}
 =h_{\A\otimes \C}\odot h_{\X}.
\end{equation}
The first equality permutes tensor factors, which leaves the curve unchanged; the second follows from \eqref{eq:cat-mixed-product} because $\X$ is classical.

By Lemma~\ref{lem:bounded-classical-brackets}, choose finite classical brackets $\X_n^\pm$ and $\Y_n^\pm$ satisfying
\[
 h_{\X_n^-}\le h_{\C}\le h_{\X_n^+},\qquad
 h_{\Y_n^-}\le h_{\A\otimes \C}\le h_{\Y_n^+},
\]
with both bracket sequences converging uniformly to their middle curves within fixed upper-threshold sectors. Using \eqref{eq:cat-exchange}, order preservation of $\odot$, and \eqref{eq:cat-mixed-product}, we obtain
\ba\label{eq:cat-bracket-tensor-order}
 h_{(\A\otimes \X_n^+)\otimes \C}
 &=h_{\A\otimes \C}\odot h_{\X_n^+}\ge h_{\A\otimes \C}\odot h_{\C}\\
 &\ge h_{\Y_n^-}\odot h_{\C}=h_{\Y_n^-\otimes \C},\\[3pt]
 h_{\Y_n^+\otimes \C}
 &=h_{\Y_n^+}\odot h_{\C}\ge h_{\A\otimes \C}\odot h_{\C}\\
 &\ge h_{\A\otimes \C}\odot h_{\X_n^-}=h_{(\A\otimes \X_n^-)\otimes \C}.
 \ea
Thus $\C$ is an explicit catalyst for both comparisons:
\be\label{eq:cat-bracket-comparisons}
 \A\otimes \X_n^+\cmaj \Y_n^-,\qquad
 \Y_n^+\cmaj \A\otimes \X_n^-.
\ee
Fix a finite classical background with curve $k$ and $0\le t\le1$. By the direct-sum construction in \eqref{eq:cat-mixture}, mixing this background with $\A\otimes \X_n^+$ and $\Y_n^-$, respectively, gives two pairs with curves
\be\label{eq:cat-background-curves}
 (1-t)k+t(h_{\A}\odot h_{\X_n^+}),
 \qquad (1-t)k+t h_{\Y_n^-}.
\ee
Tensoring each pair with $\C$ preserves its direct-sum blocks, so the positive part and trace are evaluated separately on each block. The background curve becomes $k\odot h_{\C}$ by \eqref{eq:cat-mixed-product}. The resulting curves therefore satisfy
\ba\label{eq:cat-background-tensor-order}
 &(1-t)(k\odot h_{\C})+t h_{(\A\otimes \X_n^+)\otimes \C}\\
 &\qquad\ge(1-t)(k\odot h_{\C})+t h_{\Y_n^-\otimes \C},
\ea
by the first comparison in \eqref{eq:cat-bracket-tensor-order}. Thus $\C$ also catalyzes the comparison between the mixtures in \eqref{eq:cat-background-curves}. Catalytic Lorenz monotonicity~\eqref{eq:cat-functional-monotonicity} gives
\be\label{eq:cat-first-background-comparison}
 \G[(1-t)k+t(h_{\A}\odot h_{\X_n^+})]
 \ge\G[(1-t)k+t h_{\Y_n^-}].
\ee
Applying the same construction to the second comparison in \eqref{eq:cat-bracket-tensor-order}, with pairs $\Y_n^+$ and $\A\otimes \X_n^-$, gives
\be\label{eq:cat-second-background-comparison}
 \G[(1-t)k+t h_{\Y_n^+}]
 \ge\G[(1-t)k+t(h_{\A}\odot h_{\X_n^-})].
\ee
The product estimate~\eqref{22} gives $h_{\A}\odot h_{\X_n^\pm}\to h_{\A}\odot h_{\C}$ uniformly, while $h_{\Y_n^\pm}\to h_{\A\otimes \C}$. All mixtures stay in a fixed upper-threshold sector. Passing to the limit in \eqref{eq:cat-first-background-comparison} and \eqref{eq:cat-second-background-comparison} by continuity of $\G$ yields both directions of \eqref{eq:cat-replacement}. Approximating any admissible background $k$ by finite classical curves in a common sector extends the equality to all such $k$. Taking $t=1$ gives \eqref{eq:cat-product-invariance}.

Every comparison in this proof uses $\C$ itself as the catalyst. In particular, when $\C$ is a full-rank qubit pair, only full-rank qubit catalysts are required.
\end{proof}

\subsection{Why these replacements suffice}\label{app:replacement-combinations}

We now return to our goal: $\G[h_0]=\G[h_1]$ for the curves in Lemma~\ref{lem:cat-reduction}. Fix admissible $h_0,h_1$ equal to $1-\gamma$ near zero, with equal $J$ and $K$ values, and put $u=h_1-h_0$. This is a Lipschitz function supported in a compact interval contained in $(0,\infty)$. Equality of the entropies \eqref{eq:cat-entropies} means precisely
\be\label{eq:cat-perturbation-moments}
 \int_0^\infty\frac{u(\gamma)}\gamma\,\d\gamma=0\qquad\text{and}\qquad
 \int_0^\infty\frac{u(\gamma)}{\gamma^2}\,\d\gamma=0\,.
\ee
We will approximate $u$ uniformly by finite combinations of the changes supplied by Lemma~\ref{lem:cat-replacement}. Pointwise monotonicity will turn these approximations into upper and lower bounds on $\G$; continuity will then give equality. We first establish the finite comparisons, and then show exactly what approximation is sufficient.

In Lemma~\ref{lem:cat-replacement}, we considered two pairs $\A,\C$ and showed that replacing $h_{\A}\odot h_{\C}$ by $h_{\A\otimes\C}$ inside a convex mixture leaves $\G$ unchanged. Denote their difference by
\be\label{eq:cat-defect}
 \Delta_{\A,\C}\eqdef h_{\A\otimes\C}-h_{\A}\odot h_{\C}.
\ee
A replacement with weight $t$ in \eqref{eq:cat-replacement} therefore changes the whole curve by $t\Delta_{\A,\C}$ without changing its $\G$ value.

For full-rank pairs $\A,\C$, applying the product-equality argument of Lemma~\ref{lem:cat-replacement} to $J$ and $K$ in \eqref{eq:cat-entropies}, and then using tensor-product additivity of relative entropy, gives
\ba\label{eq:cat-product-entropies}
 J[h_{\A}\odot h_{\C}]&=J[h_{\A\otimes\C}]=J[h_{\A}]+J[h_{\C}],\\
 K[h_{\A}\odot h_{\C}]&=K[h_{\A\otimes\C}]=K[h_{\A}]+K[h_{\C}].
\ea
To justify this application, both functionals are Lorenz monotone by \eqref{eq:cat-entropies}, and additivity of forward and reverse relative entropy makes them catalytically monotone. For $K$, choose the classical brackets in the proof with a common positive lower likelihood-ratio bound, as well as a common upper threshold. The final assertion of Lemma~\ref{lem:bounded-classical-brackets} permits this choice for full-rank pairs; the integrals in \eqref{eq:cat-entropies} are then over a fixed interval bounded away from zero, so $K$ is continuous under the uniform limits used there.
Consequently, $\Delta_{\A,\C}$ satisfies the same two zero-integral constraints as $u$ in \eqref{eq:cat-perturbation-moments}.

The following lemma combines finitely many replacements with pointwise monotonicity. We allow rescalings $\Delta_{\A,\C}(\gamma/s)$ because the qubit differences constructed in Lemma~\ref{lem:cat-local} will be concentrated near $\gamma=1$; rescaling moves them near $\gamma=s$.

\bmyl\label{lem:cat-combinations}
Let $h_0,h_1$ be admissible curves and let
\be\label{eq:cat-rescaled-combination-sum}
 v(\gamma)=\sum_{j=1}^{m}r_j\Delta_{\A_j,\C_j}(\gamma/s_j),
 \qquad r_j\in\mathbb R,\quad s_j>0.
\ee
If $h_1-h_0\ge v$ pointwise, then $\G[h_1]\ge\G[h_0]$. If $h_1-h_0\le v$, the conclusion is reversed. In particular, $h_1-h_0=v$ implies $\G[h_1]=\G[h_0]$.
\emyl

\begin{proof}
We first prove the lemma for the case that $s_j=1$ for every $j$, so
\be\label{eq:cat-combination-sum}
 v=\sum_{j=1}^{m}r_j\Delta_{\A_j,\C_j},
 \qquad h_1-h_0\ge v.
\ee
Put $q=\sum_j|r_j|$. If $q=0$, then $v=0$ and the conclusion follows from pointwise monotonicity. Assume $q>0$. To apply Lemma~\ref{lem:cat-replacement}, we introduce an auxiliary path whose convex-mixture representation contains the required product curves. Fix $0<\theta<1$ and, for $t\in[0,1]$, define
\be\label{eq:cat-reserved-path}
 \ell_t=(1-\theta)((1-t)h_0+t h_1)+\frac{\theta}{q}\left(
   \sum_{j:r_j>0}r_j(h_{\A_j}\odot h_{\C_j})
   +\sum_{j:r_j<0}|r_j|h_{\A_j\otimes\C_j}\right).
\ee
The coefficients in parentheses are nonnegative and sum to $q$, so the added mixture has total weight $\theta$ and every $\ell_t$ is admissible.

Our aim is to compare $\G[\ell_0]$ and $\G[\ell_1]$: as $\theta\to0^+$, these endpoints converge uniformly to $h_0$ and $h_1$, so continuity will give the desired comparison of $\G[h_0]$ and $\G[h_1]$. When $v=h_1-h_0$, we will obtain $\G[\ell_0]=\G[\ell_1]$ and hence $\G[h_0]=\G[h_1]$. To establish the general inequality in \eqref{eq:cat-combination-sum}, we replace small portions of the added components while preserving $\G$. Their combined change is at most the increment to the next path point, so pointwise monotonicity gives successive inequalities leading to $\G[\ell_0]\le\G[\ell_1]$. Choose a positive integer $N$ such that $(1-\theta)q/N\le\theta$. For each $j$ with $r_j\ne0$, define
\be\label{eq:cat-replacement-weights}
 a_j\eqdef\frac{\theta|r_j|}{q},
 \qquad b_j\eqdef\frac{(1-\theta)|r_j|}{N},
 \qquad 0<b_j\le a_j<1.
\ee
Here $a_j$ is the coefficient of the entire $j$th product-curve component in \eqref{eq:cat-reserved-path}, whereas $b_j$ is the coefficient of the portion we replace. We split that component into portions with coefficients $a_j-b_j$ and $b_j$, and change only the latter.

If $r_j>0$, the precise replacement is
\be\label{eq:cat-positive-replacement}
 a_j(h_{\A_j}\odot h_{\C_j})
 \;\longmapsto\;
 (a_j-b_j)(h_{\A_j}\odot h_{\C_j})+b_jh_{\A_j\otimes\C_j}.
\ee
The portion with coefficient $a_j-b_j$ stays in place, and the whole curve changes by $b_j\Delta_{\A_j,\C_j}$. If $r_j<0$, we make the reverse replacement:
\be\label{eq:cat-negative-replacement}
 a_jh_{\A_j\otimes\C_j}
 \;\longmapsto\;
 (a_j-b_j)h_{\A_j\otimes\C_j}+b_j(h_{\A_j}\odot h_{\C_j}).
\ee
This changes the curve by $-b_j\Delta_{\A_j,\C_j}$. In either case, the change is $(1-\theta)r_j\Delta_{\A_j,\C_j}/N$.

Each replacement preserves $\G$ by Lemma~\ref{lem:cat-replacement}. Indeed, all other terms, including the unchanged portion with coefficient $a_j-b_j$, form a nonnegative mixture of total weight $1-b_j$. After normalization they define an admissible background $k$, and \eqref{eq:cat-replacement} gives
\be\label{eq:cat-weighted-replacement-invariance}
 \G[(1-b_j)k+b_j(h_{\A_j}\odot h_{\C_j})]
 =\G[(1-b_j)k+b_jh_{\A_j\otimes\C_j}].
\ee
Read in the appropriate direction, this equality justifies both \eqref{eq:cat-positive-replacement} and \eqref{eq:cat-negative-replacement}.

After all components have been replaced, the resulting admissible curve is
\be\label{eq:cat-path-increment}
 \widetilde\ell_t=\ell_t+\frac{1-\theta}{N}v
 \le\ell_t+\frac{1-\theta}{N}(h_1-h_0)=\ell_{t+1/N}.
\ee
Thus $\G[\ell_t]=\G[\widetilde\ell_t]\le\G[\ell_{t+1/N}]$. At each path point, \eqref{eq:cat-reserved-path} supplies the same component weights, so we can repeat this comparison for $t=0,1/N,\ldots,(N-1)/N$. It follows that $\G[\ell_0]\le\G[\ell_1]$. As $\theta\downarrow0$, the endpoints converge uniformly to $h_0,h_1$ in a fixed upper-threshold sector. Continuity gives $\G[h_0]\le\G[h_1]$.

To include rescaling, fix $\A=(\rho,\sigma)$ and $s>0$. Append one scalar block and define
\be\label{eq:cat-rescaling-pair}
 \A_s\eqdef\left(\frac{s\rho\oplus1}{1+s},
                         \frac{\sigma\oplus s}{1+s}\right).
\ee
Both states are normalized. Evaluating the two blocks separately gives
\be\label{eq:cat-rescaled-curve}
 h_{\A_s}(\gamma)
 =\frac{s}{1+s}h_{\A}(\gamma/s)+\frac1{1+s}e(s\gamma).
\ee
Tensoring with $\C$ leaves the blocks separate. For the classical product, substitute \eqref{eq:cat-rescaled-curve} into \eqref{eq:cat-classical-product}, first for a finite classical second factor and then by uniform approximation. Since $e\odot h_{\C}=h_{\C}$, these calculations give
\ba\label{eq:cat-rescaled-products}
 h_{\A_s\otimes\C}(\gamma)
 &=\frac{s}{1+s}h_{\A\otimes\C}(\gamma/s)+\frac1{1+s}h_{\C}(s\gamma),\\
 (h_{\A_s}\odot h_{\C})(\gamma)
 &=\frac{s}{1+s}(h_{\A}\odot h_{\C})(\gamma/s)+\frac1{1+s}h_{\C}(s\gamma).
\ea
The contribution from the scalar block is identical in both expressions and cancels. Hence
\be\label{eq:cat-rescaled-defect}
 \Delta_{\A,\C}(\gamma/s)
 =\frac{1+s}{s}\Delta_{\A_s,\C}(\gamma).
\ee
Each rescaled term in \eqref{eq:cat-rescaled-combination-sum} is therefore an unscaled product difference with a real coefficient. The finite-sum proof applies, and the catalyst $\C$ is unchanged. Finally, applying the result to $h_0-h_1\ge-v$ proves the reversed inequality; equality follows by using both directions.
\end{proof}

We need a slightly stronger conclusion than the exact-equality case of Lemma~\ref{lem:cat-combinations}: rather than express $u=h_1-h_0$ exactly as one finite sum $v$, it will suffice to approximate $u$ uniformly by finite sums $v_n$. The following lemma shows that $\|v_n-u\|_\infty\to0$ implies $\G[h_0]=\G[h_1]$, provided all these changes vanish outside one fixed compact interval contained in $(0,\infty)$. This reduces our goal to a uniform approximation problem.

\bmyl\label{lem:cat-uniform-closure}
Let $h_0,h_1$ be admissible curves and put $u\eqdef h_1-h_0$. Suppose finite sums $v_n$ of the form \eqref{eq:cat-rescaled-combination-sum} satisfy
\be\label{eq:cat-approximation-goal}
 \|v_n-u\|_\infty\xrightarrow{n\to\infty}0,
 \qquad \supp v_n\cup\supp u\subseteq I
\ee
for all $n$, with one fixed compact interval $I\subset(0,\infty)$. Then $\G[h_0]=\G[h_1]$.
\emyl

\begin{proof}
Choose $R>\max\{1,\sup I\}$ and introduce the admissible curve
\be\label{eq:cat-upper-curve}
 b_R(\gamma)\eqdef(1-\gamma/R)_+.
\ee
Since $b_R\ge e$ everywhere and $b_R>e$ on $(0,R)$, the continuous difference $b_R-e$ has a positive minimum on $I$. For $0<\delta<1$, form the convex mixtures
\ba\label{eq:cat-sandwich-curves}
 h_{0,\delta}&\eqdef(1-\delta)h_0+\frac{\delta}{2}(e+b_R),\\
 h_{1,\delta}^-&\eqdef(1-\delta)h_1+\delta e,\\
 h_{1,\delta}^+&\eqdef(1-\delta)h_1+\delta b_R.
\ea
All three curves are admissible, and their differences satisfy
\be\label{eq:cat-sandwich-differences}
 h_{1,\delta}^{\pm}-h_{0,\delta}
 =(1-\delta)u\pm\frac{\delta}{2}(b_R-e).
\ee
For fixed $\delta$, uniform convergence lets us choose $n$ so that $(1-\delta)\|v_n-u\|_\infty\le(\delta/2)\min_{\gamma\in I}(b_R(\gamma)-e(\gamma))$. Equation~\eqref{eq:cat-sandwich-differences} then gives, on $I$,
\be\label{eq:cat-sandwich-order}
 h_{1,\delta}^--h_{0,\delta}
 \le(1-\delta)v_n
 \le h_{1,\delta}^+-h_{0,\delta}.
\ee
The inequalities also hold outside $I$, because $u=v_n=0$ there and $b_R-e\ge0$. Applying the two directions of Lemma~\ref{lem:cat-combinations} yields
\be\label{eq:cat-sandwich-values}
 \G[h_{1,\delta}^-]\le\G[h_{0,\delta}]\le\G[h_{1,\delta}^+].
\ee
As $\delta\downarrow0$, the middle curve converges uniformly to $h_0$ and both outer curves to $h_1$, all in a fixed upper-threshold sector. Continuity therefore proves $\G[h_0]=\G[h_1]$.

Although $h_0+v_n$ converges uniformly to $h_1$, it may fail to be convex and hence admissible: uniform approximation does not ensure convexity of the approximating functions. The construction \eqref{eq:cat-sandwich-curves} resolves this difficulty. The finite sums $v_n$ enter only through the pointwise comparisons \eqref{eq:cat-sandwich-order}, while every argument of $\G$ is an admissible convex mixture.
\end{proof}

\subsection{The qubit calculation}\label{app:qubit-calculation}

It therefore remains to approximate $u=h_1-h_0$ by finite combinations of qubit differences as in \eqref{eq:cat-approximation-goal}. The next lemma supplies this approximation using the two constraints \eqref{eq:cat-perturbation-moments}. Its small qubit differences are concentrated near $\gamma=1$; rescaling places them at different values of $\gamma$, and real coefficients combine them to approximate $u$. The proof constructs these coefficients explicitly.

\bmyl[Finite approximation by qubit differences]\label{lem:cat-local}
There is a family of full-rank qubit pairs $\C_\epsilon$ with the following property. For every real Lipschitz function $u$ supported in a compact interval contained in $(0,\infty)$ and satisfying \eqref{eq:cat-perturbation-moments}, there are finite sums
\ba\label{eq:cat-finite-approximation}
 v_n(\gamma)&=\sum_{j=1}^{m_n}r_{j,n}
 \Delta_{\C_{\epsilon_n},\C_{\epsilon_n}}(\gamma/s_{j,n}),\\
 &\qquad r_{j,n}\in\mathbb R,\quad s_{j,n}>0,\quad\epsilon_n\downarrow0,
\ea
such that $\|v_n-u\|_\infty\to0$ and all $v_n$ and $u$ are supported in one compact interval $I\subset(0,\infty)$.
\emyl

\begin{proof}
We now construct the finite approximations required by Lemma~\ref{lem:cat-local}. The goal throughout is to approximate the prescribed change $u$, whose two weighted integrals vanish. We first express $u$ as a second derivative, then show that small qubit differences reproduce this derivative after averaging over rescalings. Ordinary Riemann sums will give the required finite combinations.

Fix $u$ as in the lemma, and choose $0<a<b<\infty$ with $u$ supported in $[a,b]$. Define
\be\label{eq:cat-double-integral}
 f(\gamma)\eqdef\int_0^\gamma(\gamma-s)\frac{u(s)}{s^2}\,\d s,
 \qquad\gamma^2f''(\gamma)=u(\gamma).
\ee
The function $f$ is also supported in $[a,b]$. For $\gamma<a$ its defining integral is zero; for $\gamma>b$ it equals
\[
 \gamma\int_a^b\frac{u(s)}{s^2}\,\d s
 -\int_a^b\frac{u(s)}s\,\d s=0
\]
by \eqref{eq:cat-perturbation-moments}. Moreover, $f$ is twice continuously differentiable and $f''=u/\gamma^2$ is Lipschitz, since $u$ is Lipschitz and vanishes near zero. Thus second-order Taylor expansion of $f$ has a remainder bounded by a constant times the cube of the increment.

To reproduce $\gamma^2f''(\gamma)$ using product differences, take $\C_\epsilon=(\rho_\epsilon,\sigma)$ with
\be\label{eq:cat-fixed-reference}
 \sigma=\begin{pmatrix}3/4&0\\0&1/4\end{pmatrix},\qquad
 \rho_\epsilon=\begin{pmatrix}3/4&\epsilon/2\\\epsilon/2&1/4\end{pmatrix},
 \qquad0<\epsilon<\sqrt3/2,
\ee
and write $\Delta_\epsilon\eqdef\Delta_{\C_\epsilon,\C_\epsilon}$. Both states are full rank. As $\epsilon\to0$, the hypotheses approach each other, and $\Delta_\epsilon$ is confined to an interval shrinking to $\gamma=1$, as justified by the support bound below; see \eqref{eq:cat-qubit-support}. We need its normalization and its first few moments to identify the effect of averaging its rescalings.

For the normalization, define $Q_2[h]=2\int_0^\infty h(\gamma)\,\d\gamma$. Here $Q_2[h_{\rho,\sigma}]-1$ is the Bogoliubov--Kubo--Mori (BKM) $\chi^2$-divergence associated with the BKM metric; see~\cite[Sec.~2, Example~7]{PetzGhinea2011} for the metric and its name, and~\cite{Temme2010} for the family of quantum $\chi^2$-divergences. Hirche and Tomamichel~\cite[Theorem~2.8 and Eqs.~(2.61)--(2.64)]{HT2024} show that the order-two Hellinger divergence in their integral family coincides with this established divergence and satisfies
\be\label{eq:cat-chi-square-hessian}
 Q_2[h_{\rho,\sigma}]-1
 =\chi^2_{\mathrm{BKM}}(\rho\|\sigma)
 =\left.\frac{\d^2}{\d t^2}
 \bigl[(\ln2)D(\sigma+t(\rho-\sigma)\|\sigma)\bigr]\right|_{t=0}.
\ee
The Hessian calculation goes back to Petz and Ruskai~\cite{PetzRuskai1998}. The factor $\ln2$ converts our relative entropy in bits to the natural-logarithm convention. The known resolvent representation~\cite[Eqs.~(2.62)--(2.64)]{HT2024}, also expressed through the logarithmic metric in~\cite[Eq.~(54)]{Temme2010}, gives
\ba\label{eq:cat-q2-spectral}
 Q_2[h_{\rho,\sigma}]
 &=\int_0^\infty\tr\bigl[\rho(\sigma+r\mathbf1)^{-1}
                   \rho(\sigma+r\mathbf1)^{-1}\bigr]\,\d r\\
 &=\sum_{x,y}|\rho_{xy}|^2\frac{\ln s_x-\ln s_y}{s_x-s_y}.
\ea
Here $\mathbf1$ is the identity operator, $s_x>0$ are the eigenvalues of $\sigma$, and $\rho_{xy}$ are the matrix entries of $\rho$ in the same eigenbasis; the quotient is interpreted as $1/s_x$ when $s_x=s_y$. With $\ell=\ln3$, substitution into \eqref{eq:cat-q2-spectral} yields
\be\label{fq2}
 Q_2[h_{\C_\epsilon}]=1+\ell\epsilon^2,\qquad
 Q_2[h_{\C_\epsilon^{\otimes2}}]
 =1+2\ell\epsilon^2+(2/3+\ell/2)\epsilon^4.
\ee
Multiplicativity of $Q_2$ under $\odot$ follows from the classical R\'enyi divergence of order two. For a finite classical pair $\X=(\p,\q)$,
$Q_2[h_{\X}]=2^{D_2(\p\|\q)}$.
Since $h_{\X}\odot h_{\Y}=h_{\X\otimes\Y}$ for classical pairs, additivity of the classical R\'enyi divergence immediately gives
\be\label{eq:cat-q2-multiplicativity}
 Q_2[h\odot k]=Q_2[h]Q_2[k]
\ee
for finite classical curves $h,k$. Uniform classical approximation with common upper thresholds extends this identity to all admissible curves, using \eqref{22} and the continuity bound $|Q_2[f]-Q_2[g]|\le2T\|f-g\|_\infty$ when $f,g$ vanish for $\gamma\ge T$.

Applying \eqref{eq:cat-q2-multiplicativity} with $h=k=h_{\C_\epsilon}$ together with~\eqref{fq2} now gives
\ba\label{eq:cat-exact-area}
 a_\epsilon&\eqdef2\int_0^\infty\Delta_\epsilon(\gamma)\,\d\gamma
 =Q_2[h_{\C_\epsilon^{\otimes2}}]-Q_2[h_{\C_\epsilon}]^2
 =c\epsilon^4,\\
 c&=\tfrac23+\tfrac12\ln3-(\ln3)^2>\tfrac1{150}.
\ea
The bound follows from $1<\ln3<11/10$. In particular, this positive integral proves that $h_{\C_\epsilon^{\otimes2}}\ne h_{\C_\epsilon}\odot h_{\C_\epsilon}$ for every allowed $\epsilon>0$, giving the example announced in Section~\ref{app:catalytic-replacement}.

To prove the support bound, put $t_\epsilon\eqdef2\epsilon/\sqrt3\in(0,1)$. The matrix
\[
 \sigma^{-1/2}\rho_\epsilon\sigma^{-1/2}
 =\begin{pmatrix}1&t_\epsilon\\t_\epsilon&1\end{pmatrix}
\]
has eigenvalues $1\pm t_\epsilon$. It follows that
\ba\label{eq:cat-qubit-operator-bounds}
 (1-t_\epsilon)\sigma&\le\rho_\epsilon\le(1+t_\epsilon)\sigma,\\
 (1-t_\epsilon)^2\sigma^{\otimes2}
 &\le\rho_\epsilon^{\otimes2}\le(1+t_\epsilon)^2\sigma^{\otimes2}.
\ea
Thus the quantum product curve $h_{\C_\epsilon^{\otimes2}}(\gamma)=\tr(\rho_\epsilon^{\otimes2}-\gamma\sigma^{\otimes2})_+$ equals $1-\gamma$ for $\gamma\le(1-t_\epsilon)^2$ and zero for $\gamma\ge(1+t_\epsilon)^2$.

The single-pair curve likewise equals $1-\gamma$ for $\gamma\le1-t_\epsilon$ and zero for $\gamma\ge1+t_\epsilon$. Lemma~\ref{lem:bounded-classical-brackets} therefore gives finite classical approximations with likelihood ratios in $[1-t_\epsilon,1+t_\epsilon]$. Their tensor products have likelihood ratios in $[(1-t_\epsilon)^2,(1+t_\epsilon)^2]$. Passing to the uniform limit defining $\odot$, the classical product curve has the same initial and final pieces as the quantum product curve. Their difference $\Delta_\epsilon$ therefore vanishes outside
\be\label{eq:cat-qubit-support}
 [(1-2\epsilon/\sqrt3)^2,(1+2\epsilon/\sqrt3)^2].
\ee
Both endpoints tend to $1$, so for each fixed $\gamma\ne1$, the difference $\Delta_\epsilon(\gamma)$ is exactly zero for all sufficiently small $\epsilon$. Moreover, $\Delta_\epsilon$ is $2$-Lipschitz, being a difference of admissible curves, and vanishes at both endpoints of the interval \eqref{eq:cat-qubit-support}. Every point of that interval is within half its width $w_\epsilon\eqdef8\epsilon/\sqrt3$ of an endpoint. Hence its height satisfies $\|\Delta_\epsilon\|_\infty\le2(w_\epsilon/2)=w_\epsilon$, and
\[
 \int_0^\infty|\Delta_\epsilon(\gamma)|\,\d\gamma
 \le w_\epsilon\|\Delta_\epsilon\|_\infty
 \le w_\epsilon^2=\frac{64}{3}\epsilon^2.
\]
Put $b_j=\int_0^\infty(\gamma-1)^j\Delta_\epsilon(\gamma)\,\d\gamma$ for $j=0,1,2$; then $b_0=a_\epsilon/2$.
As noted after \eqref{eq:cat-product-entropies}, each full-rank product difference satisfies the same two constraints as those stated for $u$ in \eqref{eq:cat-perturbation-moments}. Taking $\A=\C=\C_\epsilon$ and subtracting the corresponding $J$ and $K$ integrals in \eqref{eq:cat-entropies} gives explicitly
\be\label{eq:cat-qubit-zero-moments}
 \int_0^\infty\frac{\Delta_\epsilon(\gamma)}\gamma\,\d\gamma=0,
 \qquad
 \int_0^\infty\frac{\Delta_\epsilon(\gamma)}{\gamma^2}\,\d\gamma=0.
\ee
To express \eqref{eq:cat-qubit-zero-moments} in terms of $b_0,b_1,b_2$, expand its weights about $\gamma=1$:
\ba\label{eq:cat-weight-expansions}
 \frac1\gamma&=1-(\gamma-1)+(\gamma-1)^2+R_1(\gamma),\\
 \frac1{\gamma^2}&=1-2(\gamma-1)+3(\gamma-1)^2+R_2(\gamma).
\ea
For $\gamma$ near $1$, Taylor's theorem bounds both remainders by $|R_i(\gamma)|\le C|\gamma-1|^3$, with a constant $C$ independent of $\epsilon$. On the support \eqref{eq:cat-qubit-support}, $|\gamma-1|=O(\epsilon)$. Together with the bound on $\int|\Delta_\epsilon|$, this gives, for $i=1,2$,
\be
 \left|\int_0^\infty R_i(\gamma)\Delta_\epsilon(\gamma)\,\d\gamma\right|
 \le C\sup_{\gamma\in\supp\Delta_\epsilon}|\gamma-1|^3
       \int_0^\infty|\Delta_\epsilon(\gamma)|\,\d\gamma
 =O(\epsilon^3)O(\epsilon^2)=O(\epsilon^5).
\ee
Substituting \eqref{eq:cat-weight-expansions} into the vanishing integrals \eqref{eq:cat-qubit-zero-moments}, the constant, linear and quadratic terms integrate to $b_0,b_1,b_2$, respectively. Hence
\be
 b_0-b_1+b_2=O(\epsilon^5),\qquad
 b_0-2b_1+3b_2=O(\epsilon^5).
\ee
Subtracting these relations gives $b_1=2b_2+O(\epsilon^5)$. Substitution into the first relation then yields $b_2=b_0+O(\epsilon^5)$, and consequently $b_1=2b_0+O(\epsilon^5)$.
For a twice continuously differentiable function $\varphi$ with Lipschitz second derivative near $1$, Taylor expansion gives
\[
 \varphi(r)=\varphi(1)+\varphi'(1)(r-1)
 +\tfrac12\varphi''(1)(r-1)^2+O(|r-1|^3).
\]
Integrating against $\Delta_\epsilon$ gives an $O(\epsilon^5)$ remainder by the same estimate used for \eqref{eq:cat-weight-expansions}. Using $b_0=a_\epsilon/2$, $b_1=a_\epsilon+O(\epsilon^5)$ and $b_2=a_\epsilon/2+O(\epsilon^5)$, we obtain
\ba
 \int\varphi(r)\Delta_\epsilon(r)\,\d r
 &=\varphi(1)b_0+\varphi'(1)b_1+\tfrac12\varphi''(1)b_2+O(\epsilon^5)\\
 &=\frac{a_\epsilon}{4}
 \bigl(2\varphi(1)+4\varphi'(1)+\varphi''(1)\bigr)+O(\epsilon^5).
\ea
Multiplying by $4/a_\epsilon$ therefore yields
\be\label{eq:cat-test-limit}
 \frac4{a_\epsilon}\int\varphi(r)\Delta_\epsilon(r)\,\d r
 =\varphi''(1)+4\varphi'(1)+2\varphi(1)+O(\epsilon).
\ee
The normalized remainder is $O(\epsilon^5)/a_\epsilon=O(\epsilon)$, since $a_\epsilon=c\epsilon^4$.

We now apply this calculation to the function $f$ chosen in \eqref{eq:cat-double-integral}. Consider the average
\be\label{eq:cat-local}
 u_\epsilon(\gamma)\eqdef\frac4{a_\epsilon}
 \int_a^b\frac{f(s)}s\Delta_\epsilon(\gamma/s)\,\d s.
\ee
For fixed $\gamma>0$, substitute $r=\gamma/s$, so $\d s/s=-\d r/r$. Since $f$ vanishes outside $[a,b]$, reversing the integration limits and then extending them gives
\ba\label{eq:cat-test-function-substitution}
 u_\epsilon(\gamma)
 &=\frac4{a_\epsilon}\int_{\gamma/b}^{\gamma/a}
       \frac{f(\gamma/r)}r\Delta_\epsilon(r)\,\d r\\
 &=\frac4{a_\epsilon}\int_0^\infty
       \varphi_\gamma(r)\Delta_\epsilon(r)\,\d r,
 \qquad \varphi_\gamma(r)\eqdef\frac{f(\gamma/r)}r.
\ea
This is the form required by \eqref{eq:cat-test-limit}. Differentiating with respect to $r$, with $\gamma$ held fixed, the product and chain rules give
\ba\label{eq:cat-test-function-derivatives}
 \varphi_\gamma'(r)
 &=-\frac{f(\gamma/r)}{r^2}-\frac{\gamma f'(\gamma/r)}{r^3},\\
 \varphi_\gamma''(r)
 &=\frac{2f(\gamma/r)}{r^3}+\frac{4\gamma f'(\gamma/r)}{r^4}
   +\frac{\gamma^2f''(\gamma/r)}{r^5},
\ea
where each differentiation of $f(\gamma/r)$ introduces the factor $-\gamma/r^2$. Evaluating at $r=1$ yields
\ba
 \varphi_\gamma'(1)&=-f(\gamma)-\gamma f'(\gamma),\\
 \varphi_\gamma''(1)&=2f(\gamma)+4\gamma f'(\gamma)+\gamma^2f''(\gamma).
\ea
Since $\varphi_\gamma(1)=f(\gamma)$, substituting these values into \eqref{eq:cat-test-limit} gives
\ba\label{eq:cat-test-function-cancellation}
 u_\epsilon(\gamma)
 &=\varphi_\gamma''(1)+4\varphi_\gamma'(1)+2\varphi_\gamma(1)+O(\epsilon)\\
 &=\bigl[2f(\gamma)+4\gamma f'(\gamma)+\gamma^2f''(\gamma)\bigr]
   -4\bigl[f(\gamma)+\gamma f'(\gamma)\bigr]+2f(\gamma)+O(\epsilon)\\
 &=\gamma^2f''(\gamma)+O(\epsilon)
 =u(\gamma)+O(\epsilon),
\ea
where the last equality uses \eqref{eq:cat-double-integral}.

By \eqref{eq:cat-qubit-support} and \eqref{eq:cat-local}, both $u_\epsilon$ and $u$ vanish outside $I=[a/2,2b]$ for sufficiently small $\epsilon$. Since $f''$ is Lipschitz and $I$ is fixed and compact, the Taylor remainder in \eqref{eq:cat-test-function-cancellation} is uniform in $\gamma\in I$. Consequently,
\be\label{eq:cat-uniform-local-limit}
 \|u_\epsilon-u\|_\infty=O(\epsilon)\longrightarrow0.
\ee

It remains to turn this average into a finite sum. Choose $\epsilon_n\downarrow0$ small enough for the stated support bound. For each fixed $\epsilon_n$, the integrand in \eqref{eq:cat-local} is jointly continuous in $(\gamma,s)\in I\times[a,b]$. Choose a sufficiently fine partition of $[a,b]$ into $m_n$ intervals. Let $s_{j,n}$ be the sample point in its $j$th interval and $\delta_{j,n}>0$ that interval's length. The partition can be chosen so that the Riemann sum
\be\label{eq:cat-finite-riemann-sum}
 v_n(\gamma)\eqdef\frac4{a_{\epsilon_n}}
 \sum_{j=1}^{m_n}\frac{f(s_{j,n})}{s_{j,n}}\delta_{j,n}
 \Delta_{\epsilon_n}(\gamma/s_{j,n})
\ee
satisfies $\|v_n-u_{\epsilon_n}\|_\infty\le1/n$. Each summand also vanishes outside $I$, so the bound holds on the whole domain. Equation~\eqref{eq:cat-finite-riemann-sum} has exactly the required form \eqref{eq:cat-finite-approximation}, with
\be
 r_{j,n}=\frac{4f(s_{j,n})\delta_{j,n}}{a_{\epsilon_n}s_{j,n}}.
\ee
Finally,
\[
 \|v_n-u\|_\infty
 \le\frac1n+\|u_{\epsilon_n}-u\|_\infty\xrightarrow{n\to\infty}0
\]
by \eqref{eq:cat-uniform-local-limit}. This proves Lemma~\ref{lem:cat-local} and completes the approximation required for \eqref{eq:cat-sandwich-values}.
\end{proof}

Applied to $u=h_1-h_0$, which is Lipschitz, Lemma~\ref{lem:cat-local} supplies exactly \eqref{eq:cat-approximation-goal}. Lemma~\ref{lem:cat-uniform-closure} then proves \eqref{eq:cat-levelsets}, and Lemma~\ref{lem:cat-reduction} gives the desired representation $\G[h]=g(J[h])$.

In applying Lemmas~\ref{lem:cat-replacement} and~\ref{lem:cat-combinations}, only the full-rank qubit catalysts $\C=\C_{\epsilon_n}$ are needed; the rescaling construction \eqref{eq:cat-rescaling-pair} changes only $\A$. An uninformative full-rank qubit supplies ordinary Lorenz monotonicity. This proves the catalyst restriction. Conversely, Umegaki's additivity cancels the finite catalyst contribution, so every continuous nondecreasing function of it satisfies the assumptions.

\section{Classical catalysis: R\'enyi monotonicity and its limits}
\label{app:classical-catalysis}

Here classical catalysts may assist arbitrary quantum source and target pairs. We prove Proposition~\ref{lem:cat-classical-additivity} and give the explicit pairs and verification for Example~\ref{ex:cat-ht-counterexample}, then explain the restrictions on catalysts and endpoints in the main theorem.

\begin{proof}[Proof of Proposition~\ref{lem:cat-classical-additivity}]
Choose finite classical pairs $\X_n$ with $h_{\X_n}\to h_{\A}$ uniformly in a fixed upper-threshold sector. The product estimate and \eqref{eq:cat-mixed-product} give
\be
 h_{\X_n\otimes \Z}=h_{\X_n}\odot h_{\Z}
 \longrightarrow h_{\A}\odot h_{\Z}=h_{\A\otimes \Z}.
\ee
All product curves remain in a common bounded sector. Classical additivity and continuity of the induced curve functional therefore yield
\ba
 \D(\A\otimes \Z)
 &=\lim_{n\to\infty}\D(\X_n\otimes \Z)\\
 &=\lim_{n\to\infty}\bigl(\D(\X_n)+\D(\Z)\bigr)
 =\D(\A)+\D(\Z).
\ea
This uses the continuous curve extension of the classical restriction, without assuming quantum continuity separately.

If $\A\otimes \Z\succ_L \B\otimes \Z$ with $\Z$ classical, Lorenz monotonicity and \eqref{eq:cat-classical-additivity} imply
\be
 \D(\A)+\D(\Z)\ge\D(\B)+\D(\Z).
\ee
Cancelling the finite catalyst value gives $\D(\A)\ge\D(\B)$. For classical endpoints $\X,\Y$ and a quantum catalyst $\C$, each tensor product still has a classical factor, so the same additivity and cancellation apply.
\end{proof}

\begin{proof}[Verification of Example~\ref{ex:cat-ht-counterexample}]
Take
\be\label{eq:cat-smooth-f}
 f(t)=\sqrt{(t-1)^2+\tfrac1{16}}-\tfrac14,
 \qquad
 f''(t)=\frac{1/16}{\bigl((t-1)^2+1/16\bigr)^{3/2}}>0.
\ee
The HT divergence~\cite{HT2024}, expressed in our curve notation, is
\be\label{eq:cat-smooth-f-functional}
 D_f(\A)=\int_0^\infty f''(\gamma)
       \bigl(h_{\A}(\gamma)-e(\gamma)\bigr)\,\d\gamma.
\ee
The positive weight makes $D_f$ Lorenz monotone. Moreover,
$\int_0^\infty f''(\gamma)\,\d\gamma=1+4/\sqrt{17}$, so
\be
 |D_f(\A)-D_f(\B)|
 \le(1+4/\sqrt{17})\|h_{\A}-h_{\B}\|_\infty.
\ee
In particular, it is finite and Lorenz continuous.

Let $\mathbf u_d$ denote the uniform distribution on $d$ outcomes. Using the distributions from the Jonathan--Plenio catalysis example~\cite{JonathanPlenio1999}, set
\ba\label{eq:cat-classical-trumping-example}
 \X&=(\p,\mathbf u_4),&\quad\p&=(1/2,1/4,1/4,0),\\
 \Y&=(\q,\mathbf u_4),&\quad\q&=(2/5,2/5,1/10,1/10),\\
 \Z&=(\mathbf r,\mathbf u_2),&\quad\mathbf r&=(3/5,2/5).
\ea
The product distributions in decreasing order are
\ba
 (\p\otimes\mathbf r)^\downarrow
 &=(.30,.20,.15,.15,.10,.10,0,0),\\
 (\q\otimes\mathbf r)^\downarrow
 &=(.24,.24,.16,.16,.06,.06,.04,.04).
\ea
The differences of their cumulative sums, multiplied by $100$, are $(6,2,1,0,4,8,4,0)$. Hence the first vector majorizes the second. Both reference distributions are $\mathbf u_8$, so $\X\otimes \Z\succ_L \Y\otimes \Z$.

On classical pairs the HT divergence reduces to $\sum_xq_xf(p_x/q_x)$. Therefore
\ba
 D_f(\X)&=\tfrac14\bigl(f(2)+2f(1)+f(0)\bigr)
       =\frac{\sqrt{17}-1}{8},\\
 D_f(\Y)&=\tfrac12\bigl(f(8/5)+f(2/5)\bigr)=\frac25.
\ea
Since $(\sqrt{17}-1)/8<2/5$, the required monotonicity is reversed. The same pairs also reverse the simpler QLD $E_1$: $h_{\X}(1)=1/4<3/10=h_{\Y}(1)$.
\end{proof}

\paragraph{Why classical approximation does not replace a quantum catalyst.}

Classical approximation cannot replace a quantum catalyst in the uniqueness proof. If $h_{\Z_n}\to h_{\C}$ uniformly for classical pairs $\Z_n$, then
\be
 h_{\A\otimes \Z_n}=h_{\A}\odot h_{\Z_n}
 \longrightarrow h_{\A}\odot h_{\C},
\ee
which need not equal $h_{\A\otimes \C}$. Approximating the catalyst's individual curve therefore need not reproduce its effect on a quantum pair.

\paragraph{Which source and target pairs are needed?}
The theorem is stated for arbitrary finite-dimensional quantum pairs, including commuting ones; it does not require every pair to be noncommuting. The proof of Lemma~\ref{lem:cat-replacement} uses the comparisons
\be
 \A\otimes \X_n^+\cmaj \Y_n^-,\qquad
 \Y_n^+\cmaj \A\otimes \X_n^-.
\ee
Here $\X_n^\pm,\Y_n^\pm$ are classical, while $\A$ is generally quantum and the catalyst is $\C$. Thus one endpoint is classical and the other generally quantum, with their roles reversed in the second comparison. Adding the same classical direct-sum background preserves this feature. In the decisive application, $\C=\C_\epsilon$ is the noncommuting qubit pair in \eqref{eq:cat-fixed-reference}.

The initial passage from $\D$ to $\G$ also needs only comparisons with a classical endpoint: for any pair $\A$, classical brackets $\X_n^-\prec_L \A\prec_L \X_n^+$ give
\be
 \D(\X_n^-)\le\D(\A)\le\D(\X_n^+),
\ee
using an uninformative catalyst, and the two bounds converge to $\G[h_{\A}]$. Thus no comparison with two noncommuting endpoints is needed. The functional remains defined on all pairs, but monotonicity need only be imposed on comparisons with at least one classical endpoint, including comparisons between two classical pairs. Proposition~\ref{lem:cat-classical-additivity} leaves every finite positive R\'enyi order compatible with classical catalysts, or with quantum catalysts when both endpoints are classical. These distinct divergences show why either restriction would lose uniqueness.

\section{Equivalent Parametrizations of the Lorenz Preorder}
\label{supp:equiv-lorenz}

Let $\rho,\sigma\in\md(A)$ and set $d\eqdef|A|$. We use the testing region defined in~\eqref{eq:main-testing-region}.
The region $\mT(\rho,\sigma)$
is compact and convex, contains $(0,0)$ and $(1,1)$, and is symmetric under
\be
(x,y)\longmapsto(1-x,1-y),
\ee
which corresponds to replacing $\Lambda$ by $I-\Lambda$. Thus its lower
boundary is determined by its upper boundary. Its upper boundary is known as the quantum Lorenz curve:
\be
\rev{\mL_{\rho,\sigma}(r)\eqdef\sup\Big\{\tr[\Lambda\rho]\,:\,0\le \Lambda\le I,\ 
\tr[\Lambda\sigma]= r
\Big\}\qquad\forall\,r\in[0,1]\,.}
\ee

It is sometimes useful to pass instead
to error coordinates
\be
\alpha(\Lambda)=1-\tr[\Lambda\rho],
\qquad
\beta(\Lambda)=\tr[\Lambda\sigma].
\ee
In these coordinates the lower boundary of the testing region is the
hypothesis-testing tradeoff function
\be
\beta_{\rho|\sigma}(\varepsilon)
\eqdef
\inf\left\{
\tr[\Lambda\sigma]:
0\le \Lambda\le I,\ 
\tr[\Lambda\rho]\ge 1-\varepsilon
\right\}.
\ee
Thus Lorenz majorization is equivalently the pointwise comparison
\be
(\rho,\sigma)\succ_L(\rho',\sigma')
\quad\Longleftrightarrow\quad
\beta_{\rho|\sigma}(\varepsilon)
\le
\beta_{\rho'|\sigma'}(\varepsilon)
\qquad
\forall\,\varepsilon\in[0,1].
\ee
Equivalently, in terms of the hypothesis-testing divergence,
\be
D_H^\varepsilon(\rho\|\sigma)
\eqdef
-\log \beta_{\rho|\sigma}(\varepsilon),
\ee
one has
\be
(\rho,\sigma)\succ_L(\rho',\sigma')
\quad\Longleftrightarrow\quad
D_H^\varepsilon(\rho\|\sigma)
\ge
D_H^\varepsilon(\rho'\|\sigma')
\qquad
\forall\,\varepsilon\in[0,1].
\ee

The hockey-stick divergences used in the main text are related to this
tradeoff curve by convex duality. Indeed,
\be
E_\gamma(\rho\|\sigma)
=
\sup_{0\le \Lambda\le I}
\left\{
1-\alpha(\Lambda)-\gamma\beta(\Lambda)
\right\}
=
\sup_{0\le\varepsilon\le1}
\left\{
1-\varepsilon-\gamma\beta_{\rho|\sigma}(\varepsilon)
\right\}.
\ee
Conversely (see~\cite{BG2017}),
\be
\beta_{\rho|\sigma}(\varepsilon)
=
\sup_{\gamma>0}
\frac{1-\varepsilon-E_\gamma(\rho\|\sigma)}{\gamma}.
\ee
Thus the hypothesis-testing tradeoff curve and the hockey-stick family contain
the same information, and both determine the Lorenz preorder.

Another equivalent scalar parametrization is given by the Hilbert
$\alpha$-divergences of~\cite{BG2017}. For $\alpha>1$, define
\be
\sup{}_{\!\alpha}(\rho/\sigma)
\eqdef
\sup_{\alpha^{-1}I\le \Lambda\le I}
\frac{\tr[\Lambda\rho]}{\tr[\Lambda\sigma]}\,,
\ee
and
\be
H_\alpha(\rho\|\sigma)
\eqdef
\frac{\alpha}{\alpha-1}
\log_2 \sup{}_{\!\alpha}(\rho/\sigma)\,.
\ee
Theorem~2 of~\cite{BG2017} shows that
\be
(\rho,\sigma)\succ_L(\rho',\sigma')
\quad\Longleftrightarrow\quad
\begin{cases}
H_\alpha(\rho\|\sigma)
\ge
H_\alpha(\rho'\|\sigma'),\\[0.3em]
H_\alpha(\sigma\|\rho)
\ge
H_\alpha(\sigma'\|\rho'),
\end{cases}
\qquad
\forall\,\alpha>1.
\ee
Hence the testing region, the hypothesis-testing tradeoff curve, the
hockey-stick divergences, and the Hilbert $\alpha$-divergences are equivalent
parametrizations of the same Lorenz preorder.

\section{Geometry of Quantum Testing Regions}
\label{supp:testing-geometry}

A useful description of the testing region comes from the extreme effects.
The extreme points of the operator interval $0\le \Lambda \le I$ are precisely
the orthogonal projections. For $k=0,\ldots,d$, define
\be
\mW_k(\rho,\sigma)
\coloneqq
\left\{
\bigl(\tr[P\sigma],\tr[P\rho]\bigr):
P=P^2=P^\ast,\ \tr P=k
\right\}.
\ee
Thus $\mW_k(\rho,\sigma)$ is obtained by taking all rank-$k$ projectors $P$
and recording the two probabilities $\tr[P\sigma]$ and $\tr[P\rho]$. In other
words, it is the image, in the plane, of the set of all $k$-dimensional
subspaces of the Hilbert space. Mathematically, this set is the complex
Grassmannian,
\be
\mathrm{Proj}_k(\mathbb C^d)
\cong
\mathrm{Gr}(k,d)
\cong
\frac{U(d)}{U(k)\times U(d-k)}.
\ee
This geometric identification is not needed in what follows, but it is useful
to keep in mind: each rank-$k$ projector contributes one point to the plane,
and $\mW_k(\rho,\sigma)$ is the resulting ``shadow'' of the family of such
projectors.

Since $\mT(\rho,\sigma)$ is the image of the convex set $0\le \Lambda\le I$
under the linear map
\be
\Lambda
\longmapsto
\bigl(\tr[\Lambda\sigma],\tr[\Lambda\rho]\bigr),
\ee
it is the convex hull of the points obtained from the extreme effects, namely
\be
\mT(\rho,\sigma)
=
\operatorname{conv}
\left(
\bigcup_{k=0}^d
\mW_k(\rho,\sigma)
\right).
\ee
Thus the quantum Lorenz curve is the upper boundary, or upper envelope, of the
convex hull of these projection shadows.

\begin{figure}[h]
\includegraphics[width=0.3\columnwidth]{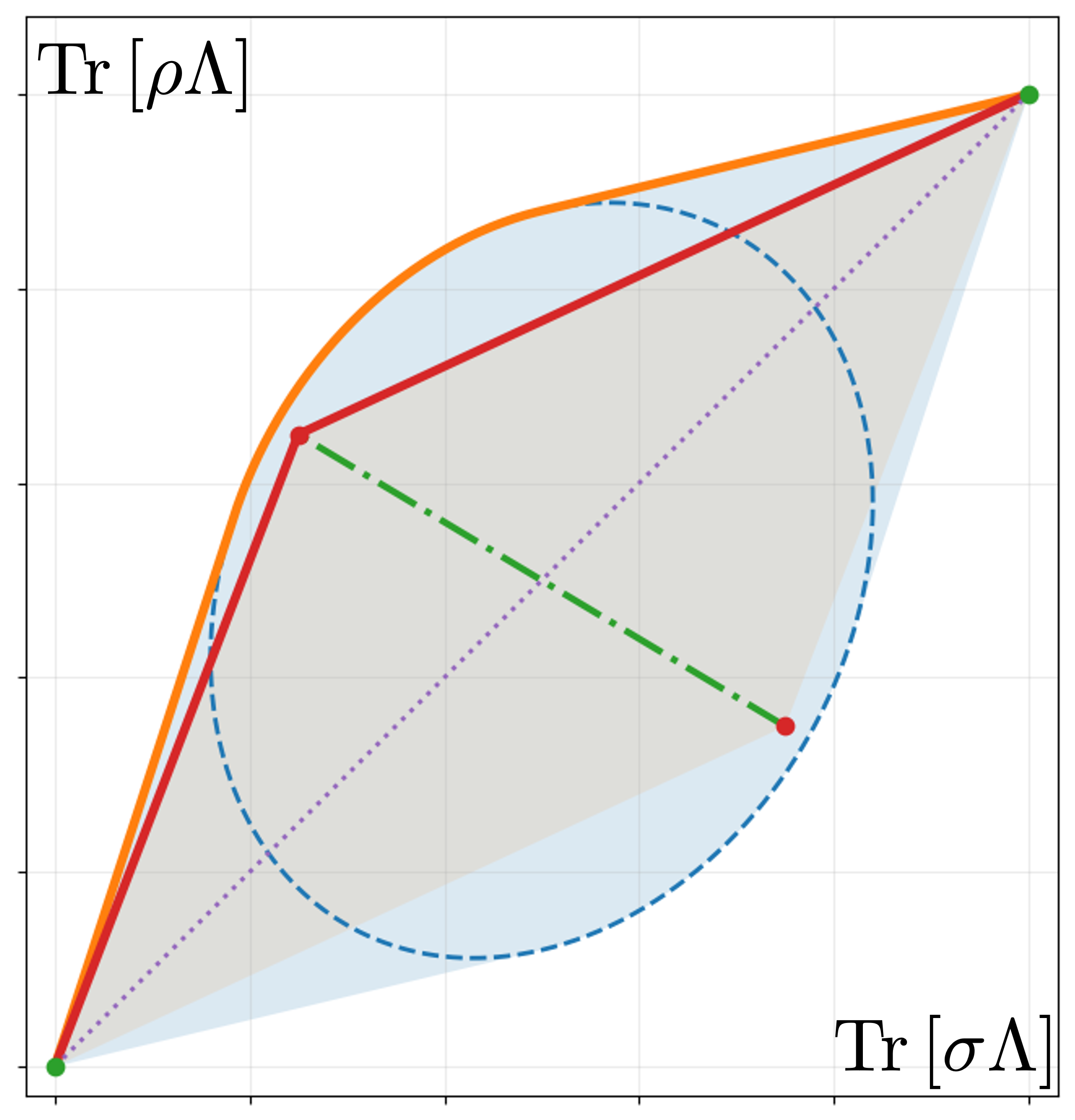}
\caption{Testing regions for qubit pairs. The orange boundary corresponds to
a genuinely quantum pair, while the red boundary corresponds to a commuting,
hence classical, pair.}
\label{ellipse}
\end{figure}

\subsection{Qubit Example}

For a qubit, $d=2$, the only nontrivial projectors have rank one. These are
the usual pure-state projectors, or equivalently the sharp yes/no measurements
associated with directions on the Bloch sphere. Thus we can write
\be
P_{\mathbf n}
=
\frac12(I+\mathbf n\cdot\boldsymbol\tau),
\qquad
|\mathbf n|=1,
\ee
where $\boldsymbol\tau$ denotes the vector of Pauli matrices. In this way, the
set of rank-one projectors is simply the Bloch sphere. Mathematically, this is
the identification
$
\mathrm{Proj}_1(\mathbb C^2)
\cong
\mathbb CP^1
\cong
S^2 
$.

Writing the two states as
\be
\rho=\frac12(I+\mathbf r\cdot\boldsymbol\tau),
\qquad
\sigma=\frac12(I+\mathbf s\cdot\boldsymbol\tau),
\ee
we find that each measurement direction $\mathbf n$ gives the point
\be
\bigl(\tr[P_{\mathbf n}\sigma],\tr[P_{\mathbf n}\rho]\bigr)
=
\left(
\frac12(1+\mathbf n\cdot\mathbf s),
\frac12(1+\mathbf n\cdot\mathbf r)
\right)
\ee
in the testing plane. Thus $\mW_1(\rho,\sigma)$ is obtained by looking at the
Bloch sphere only through the two expectation values
$\mathbf n\cdot\mathbf s$ and $\mathbf n\cdot\mathbf r$. Geometrically, this
is a two-dimensional shadow of the Bloch sphere.

The result is a filled ellipse in the plane, possibly degenerate to a line
segment or a point. Denoting this ellipse by $E=\mW_1(\rho,\sigma)$, the full
testing region is
\be
\mT(\rho,\sigma)
=
\operatorname{conv}\big\{(0,0),(1,1), E\big\}\,.
\ee
Here $(0,0)$ and $(1,1)$ come from the trivial effects $\Lambda=0$ and
$\Lambda=I$, while $E$ comes from all rank-one projective measurements.

Therefore, for qubits, the quantum Lorenz curve is the upper boundary of the
convex hull of this ellipse together with the two trivial points. If $\rho$
and $\sigma$ commute, their Bloch vectors $\mathbf r$ and $\mathbf s$ are
collinear. In that case the ellipse collapses to a line segment, and the upper
boundary reduces to the usual classical binary Lorenz polygon; see
Fig.~\ref{ellipse}.

\subsection{The General Case}

For $d>2$, the correct generalization of the qubit picture is not simply
``more ellipses.'' Instead, one must consider sharp yes/no measurements of all
possible ranks. For each $k=1,\ldots,d-1$, the set
\be
\mW_k(\rho,\sigma)
=
\left\{
\bigl(\tr[P\sigma],\tr[P\rho]\bigr):
P=P^2=P^\ast,\ \tr P=k
\right\}
\ee
is obtained by taking all rank-$k$ projectors $P$ and plotting the two
probabilities $\tr[P\sigma]$ and $\tr[P\rho]$ in the testing plane. These are
the rank-$k$ projection shadows. Mathematically, the family of rank-$k$
projectors is a Grassmannian, but for the present purpose it is enough to
think of it as the set of all $k$-dimensional sharp measurement outcomes.

Unlike the qubit case, these shadows are generally not ellipses. Nevertheless,
their supporting lines are still controlled by spectra. To see this, consider
an arbitrary real direction $(a,b)$ in the testing plane. The point of
$\mW_k$ farthest in this direction is obtained by maximizing
\be
a\,\tr[P\sigma]+b\,\tr[P\rho]
=
\tr[P(a\sigma+b\rho)]
\ee
over all rank-$k$ projectors. By Ky Fan's maximum principle,
\be
h_{\mW_k}(a,b)
=
\max_{\tr P=k}
\tr[P(a\sigma+b\rho)]
=
\sum_{j=1}^k
\lambda_j^\downarrow(a\sigma+b\rho)\eqdef\|a\sigma+b\rho\|_{(k)}\,,
\ee
where $\lambda_1^\downarrow,\ldots,\lambda_d^\downarrow$ are the eigenvalues
in decreasing order.

\begin{figure}[h]
\includegraphics[width=0.5\columnwidth]{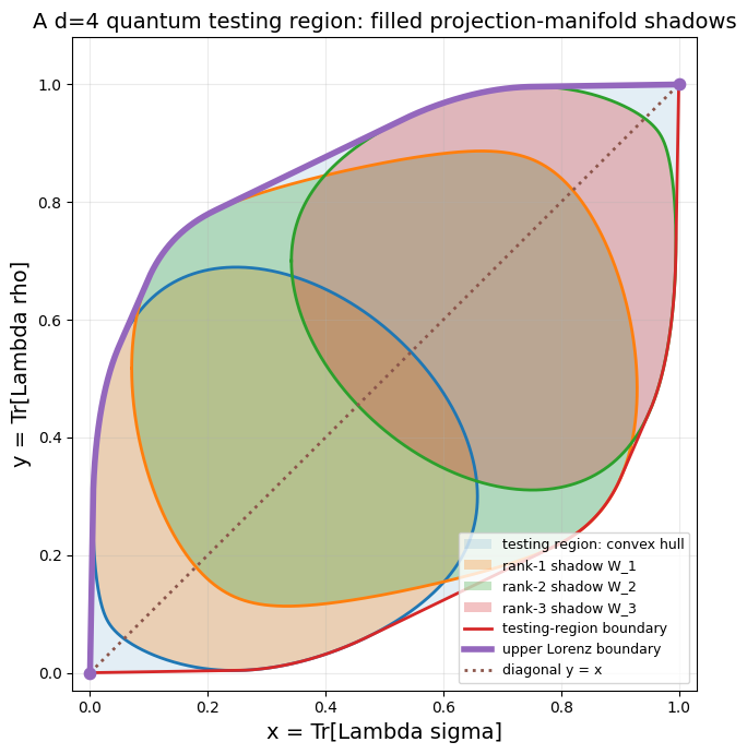}
\caption{
Testing region for a $d=4$ quantum pair. The three colored sets are the
rank-$1$, rank-$2$, and rank-$3$ projection shadows
$\mW_1$, $\mW_2$, and $\mW_3$. Taking their convex hull, together with the
trivial points $(0,0)$ and $(1,1)$ coming from $\Lambda=0$ and $\Lambda=I$,
gives the full testing region.
}
\label{grassmann}
\end{figure}

For the upper Lorenz boundary, the relevant supporting directions have the
form $(-\gamma,1)$ with $\gamma\ge 0$. Therefore the corresponding Hermitian
pencil is
$
\rho-\gamma\sigma 
$.
In this direction one obtains
\be
h_{\mW_k}(-\gamma,1)
=
\|\rho-\gamma\sigma\|_{(k)}\,.
\ee
For the full testing region, one allows all effects $0\le \Lambda\le I$, not
only projectors. Hence
\be
\max_{0\le\Lambda\le I}
\tr[\Lambda(\rho-\gamma\sigma)]
=
\tr(\rho-\gamma\sigma)_+=E_\gamma(\rho\|\sigma)\, .
\ee
In other
words, the optimal effect is the projector onto the positive eigenspace of
$\rho-\gamma\sigma$. See Fig.~\ref{grassmann}.

We finally note that finite-dimensional quantum comparisons retain a spectral
analogue of the classical vertex test, although in a less elementary form. In
the classical case, Lorenz curves are polygonal, so it is enough to compare
them at the vertices of the target Lorenz polygon. In the quantum case, the
corresponding distinguished points are spectral: in the hockey-stick
formulation,
the non-smooth points occur when eigenvalues of the pencil
$\rho-\gamma\sigma$ cross zero. Thus the quantum comparison is still governed by
finite-dimensional spectral data, but not simply by a polygonal vertex test:
one may also have to account for smooth tangencies between the relevant
hockey-stick curves.

The mixed classical-quantum cases simplify further. If the target pair is
classical, then its Lorenz curve is polygonal, so a quantum source only needs
to be compared with the target at the finitely many target vertices. If the
source pair is classical, then its hockey-stick function is piecewise linear,
with kinks at the likelihood ratios $p_x/q_x$. Since the quantum
hockey-stick function is convex in $\gamma$, the inequality can then be
checked at these finitely many source kinks, together with the endpoints.

\section{Quantum Lorenz Majorization as Convex-Order Majorization}
\label{supp:lorenz-convex-order}

Let $\rho,\sigma\in\md(A)$. The hockey-stick description also admits a layer-cake Stieltjes measure representation. Define the measure
$\d\mu_{\rho,\sigma}$ through the primitive
\be
\mu_{\rho,\sigma}(r)
\eqdef
-\tr\!\left[\sigma\{\rho>r\sigma\}\right],
\qquad r>0.
\ee
Here we record the precise relation between this measure, the hockey-stick
divergences, and convex-order majorization. The key point is that
$E_\gamma(\rho\|\sigma)$ is the stop-loss transform of
$\d\mu_{\rho,\sigma}$.

Here, $\d\mu_{\rho,\sigma}$ is the positive Stieltjes measure on
$[0,\infty)$, with the convention that its tail satisfies
\be
\int_{(r,\infty)}\d\mu_{\rho,\sigma}
=
-\mu_{\rho,\sigma}(r)\,,
\qquad r>0\,,
\ee
and with the remaining mass, if any, placed at the origin.

\bmyl[Hockey-stick divergences as stop-loss transforms]
\label{lem:hockey-stop-loss}
If $\supp(\rho)\subseteq\supp(\sigma)$, then, for every $\gamma\ge0$,
\be
E_\gamma(\rho\|\sigma)
=
\int_0^\infty
(r-\gamma)_+\,\d\mu_{\rho,\sigma}(r).
\ee
For a general pair, the corrected identity is
\be
E_\gamma(\rho\|\sigma)
=
\tr[\rho(I-\Pi_\sigma)]
+
\int_0^\infty
(r-\gamma)_+\,\d\mu_{\rho,\sigma}(r)\,,
\ee
where $\Pi_{\sigma}$ denotes the projection onto the support of $\sigma$.
\emyl

\begin{proof}
Let
\be
f_\gamma(r)\eqdef (r-\gamma)_+ \,.
\ee
We first record the Stieltjes integration-by-parts identity used below. For
$R>\gamma$, integration by parts for the positive Stieltjes measure
$\d\mu_{\rho,\sigma}$ gives
\be\label{eq:stieltjes-stop-loss-mu}
\int_{[0,R]}
f_\gamma(r)\,\d\mu_{\rho,\sigma}(r)
=
-\int_\gamma^R
\mu_{\rho,\sigma}(r)\,\d r
+
(R-\gamma)\mu_{\rho,\sigma}(R)\,.
\ee
Indeed, this is the usual identity
\be
\int_{[0,R]} f\,\d\mu
=
f(R)\mu([0,R])
-
\int_0^R \mu([0,r])\,\d f(r)\,,
\ee
applied to $f=f_\gamma$, using
$\d f_\gamma(r)=\mathbf 1_{\{r>\gamma\}}\,\d r$ and the tail relation
$\int_{(r,\infty)}\d\mu_{\rho,\sigma}=-\mu_{\rho,\sigma}(r)$. The possible atom
at the origin does not contribute, since $f_\gamma(0)=0$.

Assume first that $\supp(\rho)\subseteq\supp(\sigma)$. Since the dimension is
finite, there exists $C<\infty$ such that
$
\rho\le C\sigma
$.
Hence, for $r\ge C$ we have both
$
E_r(\rho\|\sigma)=0
$ and
$
\mu_{\rho,\sigma}(r)=0
$.
Taking $R\ge C$ in \eqref{eq:stieltjes-stop-loss-mu} gives
\be\label{eq:stop-loss-tail-identity-mu}
\int_0^\infty
(r-\gamma)_+\,\d\mu_{\rho,\sigma}(r)
=
-\int_\gamma^\infty
\mu_{\rho,\sigma}(r)\,\d r.
\ee

It remains to identify the last integral with the hockey-stick divergence. Consider the function
\be\label{rmap}
F(r)\eqdef
E_r(\rho\|\sigma)
=
\tr(\rho-r\sigma)_+.
\ee
We call $r>0$ regular if $0$ is not an eigenvalue of the Hermitian pencil
$\rho-r\sigma$ (restricted to $\supp(\rho+\sigma)$). Equivalently, no eigenvalue
of this pencil crosses zero at $r$. Thus, in finite dimensions the set of nonregular
points is finite. At every regular point, $F(r)$ is differentiable and the
standard derivative formula for the positive part gives (see, e.g.,~\cite{CL2025,LHC2025,CGLL2025})
\be\label{eq:hockey-stick-derivative-mu}
\frac{\d}{\d r}E_r(\rho\|\sigma)
=
-
\tr\!\left[\sigma\{\rho>r\sigma\}\right]
=
\mu_{\rho,\sigma}(r).
\ee
Moreover, $F(r)$ is continuous and piecewise continuously differentiable, hence absolutely continuous on compact intervals. Since $F(r)=0$ for all $r\ge C$,
integrating \eqref{eq:hockey-stick-derivative-mu} from $\gamma$ to $C$ gives
\be
E_\gamma(\rho\|\sigma)
=
-\int_\gamma^C
\mu_{\rho,\sigma}(r)\,\d r
=
-\int_\gamma^\infty
\mu_{\rho,\sigma}(r)\,\d r.
\ee
Combining this with \eqref{eq:stop-loss-tail-identity-mu} proves
\be
E_\gamma(\rho\|\sigma)
=
\int_0^\infty
(r-\gamma)_+\,\d\mu_{\rho,\sigma}(r)
\ee
whenever $\supp(\rho)\subseteq\supp(\sigma)$.

For a general pair, the derivative identity
\eqref{eq:hockey-stick-derivative-mu} still holds at regular points, but the
hockey-stick divergence need not vanish at infinity. Instead,
\be
\lim_{r\to\infty}
E_r(\rho\|\sigma)
=
\tr[\rho(I-\Pi_\sigma)]\,.
\ee
Therefore, integrating \eqref{eq:hockey-stick-derivative-mu} from $\gamma$ to
$R$ and sending $R\to\infty$ yields
\be
E_\gamma(\rho\|\sigma)
=
\tr[\rho(I-\Pi_\sigma)]
-
\int_\gamma^\infty
\mu_{\rho,\sigma}(r)\,\d r.
\ee
In particular, the integral on the right is finite. Since
$-\mu_{\rho,\sigma}(r)$ is nonincreasing and nonnegative, this also implies
\be
\lim_{R\to\infty}
(R-\gamma)\mu_{\rho,\sigma}(R)
=
0.
\ee
Taking the limit $R\to\infty$ in
\eqref{eq:stieltjes-stop-loss-mu} gives
\be
\int_0^\infty
(r-\gamma)_+\,\d\mu_{\rho,\sigma}(r)
=
-\int_\gamma^\infty
\mu_{\rho,\sigma}(r)\,\d r.
\ee
Consequently,
\be
E_\gamma(\rho\|\sigma)
=
\tr[\rho(I-\Pi_\sigma)]
+
\int_0^\infty
(r-\gamma)_+\,\d\mu_{\rho,\sigma}(r).
\ee
This completes the proof.
\end{proof}

We now obtain the convex-order characterization of the Lorenz preorder. Let
$\d\mu_{\rho,\sigma}$ and $\d\mu_{\rho',\sigma'}$ be the layer-cake measures
associated with the source and target pairs. Suppose $\supp(\rho)\subseteq\supp(\sigma)$ and $\supp(\rho')\subseteq\supp(\sigma')$. Then:

\bmyc
\label{cor:lorenz-convex-order}
\rev{Let $\supp(\rho)\subseteq\supp(\sigma)$ and $\supp(\rho')\subseteq\supp(\sigma')$. Then}
\be
(\rho,\sigma)\succ_L(\rho',\sigma')
\qquad\Longleftrightarrow\qquad
\d\mu_{\rho,\sigma}
\succ
\d\mu_{\rho',\sigma'}
\ee
in convex order.
\emyc

\begin{proof}
By the hockey-stick characterization of quantum Lorenz majorization,
\be
(\rho,\sigma)\succ_L(\rho',\sigma')
\qquad\Longleftrightarrow\qquad
E_\gamma(\rho\|\sigma)
\ge
E_\gamma(\rho'\|\sigma')
\qquad
\forall\,\gamma\ge0.
\ee
Using Lemma~\ref{lem:hockey-stop-loss} for both pairs, this is equivalent to
\be
\int_0^\infty
(r-\gamma)_+\,\d\mu_{\rho,\sigma}(r)
\ge
\int_0^\infty
(r-\gamma)_+\,\d\mu_{\rho',\sigma'}(r)
\qquad
\forall\,\gamma\ge0.
\ee
By the stop-loss characterization of convex order for probability measures on
$[0,\infty)$, the preceding inequalities are precisely
$
\d\mu_{\rho,\sigma}
\succ
\d\mu_{\rho',\sigma'}
$.
\end{proof}

\section{Uniform distance and brackets in the Lorenz-boundary picture}
\label{supp:curve-distance}

The main proof uses hockey-stick curves throughout. Here we relate the norm~\eqref{eq:main-uniform-norm} and the brackets of Lemma~\ref{lem:bounded-classical-brackets} to the Lorenz boundaries defined in Appendix~\ref{supp:equiv-lorenz}. The norm separates Lorenz equivalence classes: distinct labeled pairs may have the same curve. 
Let $\X=(\p,\q)$ and $\widetilde{\X}=(\tilde\p,\tilde\q)$ be two classical pairs and $\mL_{\X}$ and $\mL_{\tilde\X}$ denote the corresponding
classical Lorenz curves. The following identity shows that $\|\cdot\|_\infty$ is
equivalently the uniform distance between Lorenz curves.

\bmyl[Uniform distance in two curve parametrizations]
\label{lem:lorenz-distance-equivalence}
For finite classical pairs $\X$ and $\tilde\X$,
\be\label{lorenz-distance-equivalence}
\left\|
h_{\X}-h_{\tilde\X}
\right\|_\infty
=
\sup_{0\le r\le1}
\left|
\mL_{\X}(r)
-
\mL_{\tilde\X}(r)
\right|.
\ee
\emyl

\begin{proof}
Let the left-hand side of~\eqref{lorenz-distance-equivalence} be $\eps$
and the right-hand side be $\delta$. The duality between the Lorenz curve
and the hockey-stick curve gives
\be
h_{\X}(\gamma)
=
\sup_{0\le r\le1}
\left\{
\mL_{\X}(r)-\gamma r
\right\}.
\ee
By definition of $\delta$, $\mL_{\X}(r)\le \mL_{\tp,\tq}(r)+\delta$ for all $r\in[0,1]$. Hence, for every $\gamma\ge0$,
\be
\begin{aligned}
h_{\X}(\gamma)
&\le
\sup_{0\le r\le1}
\left\{
\mL_{\tilde\X}(r)-\gamma r+\delta
\right\}  \\
&=
h_{\tilde\X}(\gamma)+\delta .
\end{aligned}
\ee
Interchanging the two pairs gives
\be
\left|
h_{\X}(\gamma)
-
h_{\tilde\X}(\gamma)
\right|
\le
\delta
\qquad
\forall\,\gamma\ge0.
\ee
Thus $\eps\le\delta$.

Conversely, the inverse dual formula gives
\be
\mL_{\X}(r)
=
\inf_{\gamma\ge0}
\left\{
h_{\X}(\gamma)+\gamma r
\right\}.
\ee
By definition of $\eps$, $h_{\X}(\gamma)\le h_{\tilde\X}(\gamma) +\eps$ for all $\gamma\ge 0$. Therefore, for every $r\in[0,1]$,
\be
\begin{aligned}
\mL_{\X}(r)
&\le
\inf_{\gamma\ge0}
\left\{
h_{\tilde\X}(\gamma)+\eps+\gamma r
\right\}  \\
&=
\mL_{\tilde\X}(r)+\eps .
\end{aligned}
\ee
Interchanging the two pairs gives
\be
\left|
\mL_{\X}(r)
-
\mL_{\tilde\X}(r)
\right|
\le
\eps
\qquad
\forall\,r\in[0,1].
\ee
Thus $\delta\le\eps$, and the two quantities are equal.
\end{proof}

\subsection*{The classical brackets in geometric form}

\begin{figure}[h]
\includegraphics[width=0.8\columnwidth]{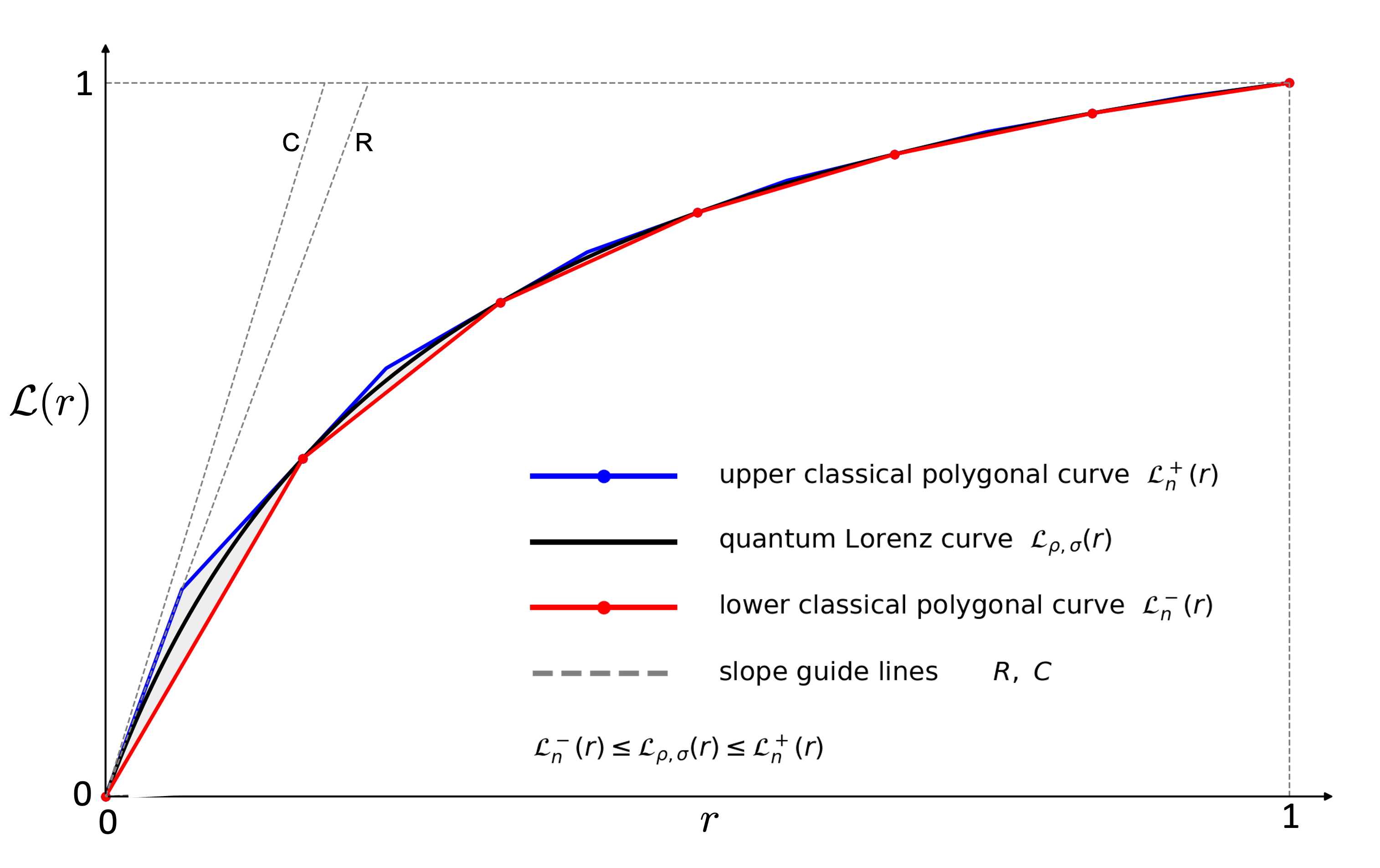}
\caption{
Classical bracketing of a quantum Lorenz curve. The lower polygonal curve is the chordal interpolation of $\mL_{\rho,\sigma}\equiv\mL_{\A}$, while the upper polygonal curve is the lower envelope of supporting lines. Both classical curves are concave, have slopes bounded by $R$, and converge uniformly to $\mL_{\A}$ as the partition mesh tends to zero.
}
\label{brackets}
\end{figure}

For $\rho\le R\sigma$, the Lorenz boundary $\mL_{\rho,\sigma}$ is concave and nondecreasing, joins $(0,0)$ to $(1,1)$, and has slopes in $[0,R]$. In these coordinates the bracketing construction uses chordal interpolation for a lower curve $\mL_n^-$ and the lower envelope of supporting lines for an upper curve $\mL_n^+$, as shown in Fig.~\ref{brackets}. At the endpoints one can use the supporting lines $y=Rr$ and $y=1$. As the partition mesh tends to zero,
\be
 \mL_n^-\le\mL_{\A}\le\mL_n^+,
 \qquad\|\mL_n^+-\mL_n^-\|_\infty\longrightarrow0.
\ee
Each polygonal curve is finite classical: for its vertices $(r_x,y_x)$, set $q_x=r_x-r_{x-1}$ and $p_x=y_x-y_{x-1}$. The endpoint values give $\sum_xp_x=\sum_xq_x=1$, while the slopes give $0\le p_x/q_x\le R$. Lemma~\ref{lem:lorenz-distance-equivalence} transfers the uniform gap to the corresponding hockey-stick curves. This gives the geometric version of the direct construction in Appendix~\ref{supp:classical-brackets}.

\end{document}